\documentclass[journal, twocolumn,final]{IEEEtran}
\usepackage{amsfonts}
\usepackage{amssymb}
\usepackage{amsmath}
\usepackage{dsfont}
\usepackage{graphicx,subfigure}
\usepackage{bm}
\usepackage{cite}
\usepackage{bigstrut}
\usepackage{float}
\usepackage{balance}
\usepackage{lipsum}
\usepackage{psfrag}
\usepackage{xcolor}
\usepackage{xfrac}
\usepackage{hyperref}
\usepackage[ruled,vlined]{algorithm2e}
\usepackage{tikz}
\usepackage{pgfplots}
\pgfplotsset{compat=1.15}
\usepackage{mathrsfs}
\usepackage{diagbox}
\usetikzlibrary{arrows}
\usetikzlibrary{decorations.pathreplacing}
\usetikzlibrary{fadings}

\newtheorem{proposition}{Proposition}
\newtheorem{remark}{Remark}

\usepackage{mathtools}
\SetKwInput{KwInput}{Input}                
\SetKwInput{KwOutput}{Output}

\begin{document}
\IEEEoverridecommandlockouts
\newcommand\norm[1]{\left\lVert#1\right\rVert}
\newcolumntype{P}[1]{>{\centering\arraybackslash}p{#1}}

\title{Waveform Design for Over-the-Air Computing}
\author{Nikos G. Evgenidis,~\IEEEmembership{Graduate Student Member,~IEEE}, Nikos A. Mitsiou,~\IEEEmembership{Graduate Student Member,~IEEE}, Sotiris A. Tegos,~\IEEEmembership{Senior Member,~IEEE}, Panagiotis D. Diamantoulakis,~\IEEEmembership{Senior Member,~IEEE}, \\ Panagiotis Sarigiannidis,~\IEEEmembership{Member,~IEEE}, Ioannis T. Rekanos,~\IEEEmembership{Senior Member,~IEEE}, \\ and George K. Karagiannidis,~\IEEEmembership{Fellow,~IEEE}
\thanks{N. G. Evgenidis, N. A. Mitsiou, S. A. Tegos, P. D. Diamantoulakis, I. T. Rekanos and G. K. Karagiannidis are with the Department of Electrical and Computer Engineering, Aristotle University of Thessaloniki, 54124 Thessaloniki, Greece (e-mails: nevgenid@ece.auth.gr, nmitsiou@auth.gr, tegosoti@auth.gr, padiaman@auth.gr, rekanos@auth.gr, geokarag@auth.gr).}
\thanks{P. Sarigiannidis is with the Department of Electrical and Computer Engineering, University of Western Macedonia, 50100 Kozani, Greece (e-mail: psarigiannidis@uowm.gr).}
\thanks{This work has received funding from the European Union’s Horizon Europe research innovation action programme under grant agreement No. 101135423 – ENACT. This work has also received funding from the National Recovery and Resilience Plan ‘Greece 2.0’ under the European's Union's NextGenerationEU framework (A$\Sigma$KO$\Sigma$ - grant agreement no. Y$\Pi$3TA-0560864). The work of N. G. Evgenidis was supported by the Hellenic Foundation for Research and Innovation (HFRI) under the 5th Call for HFRI PhD Fellowships (Fellowship Number: 21069).}
\vspace{-.5cm}
}
\maketitle

\begin{abstract}
In response to the increasing number of devices expected in next-generation networks, a shift to over-the-air (OTA) computing has been proposed. By leveraging the superposition of multiple access channels, OTA computing enables efficient resource management by supporting simultaneous uncoded transmission in the time and frequency domains. To advance the integration of OTA computing, our study presents a theoretical analysis that addresses practical issues encountered in current digital communication transceivers, such as transmitter synchronization (sync) errors and intersymbol interference (ISI). To this end, we investigate the theoretical mean squared error (MSE) for OTA transmission under sync errors and ISI, while also exploring methods for minimizing the MSE in OTA transmission. Using alternating optimization, we also derive optimal power policies for both the devices and the base station. In addition, we propose a novel deep neural network (DNN)-based approach to design waveforms that improve OTA transmission performance under sync errors and ISI. To ensure a fair comparison with existing waveforms such as raised cosine (RC) and better-than-raised-cosine (BTRC), we incorporate a custom loss function that integrates energy and bandwidth constraints along with practical design considerations such as waveform symmetry. Simulation results validate our theoretical analysis and demonstrate performance gains of the designed pulse over RC and BTRC waveforms. To facilitate testing of our results without the need to rebuild the DNN structure, we also provide curve-fitting parameters for the selected DNN-based waveforms.
\end{abstract}


\section{Introduction} \label{sec:intro}
The next generation of wireless networks is expected to enable many applications for devices in the physical layer to reduce delay. One of these is considered to be computing, a functionality currently performed in higher layers, as the latter is extremely important for many real-world scenarios, such as autonomous driving, etc. This shift from traditional communication systems to more goal-oriented tasks requires the introduction of more sophisticated techniques to manage the available resources according to the task to be performed. In this direction, and with respect to (w.r.t.) computing, over-the-air (OTA) computing has been proposed as an interesting alternative to the conventional receive-then-compute paradigm \cite{evgenidis2024multiple}. OTA computing takes advantage of the multiple access channel (MAC) superposition principle to allow simultaneous transmission of multiple devices to compute a desired target function through wireless data aggregation \cite{zhu_over--air_2021}. As a result, OTA computing can achieve better resource management as well as improved computation efficiency due to the distributed nature of the method.

On the other hand, as shown in the pioneering work \cite{Gastpar2008}, the use of uncoded analog transmission is the optimal transmission method for OTA computing. Analog transmission implies that any value can be transmitted and no conversion to bits is performed on the received signal, while the latter contains the information of the desired computation. Nevertheless, most current communication systems are based on digital transmission of symbols using appropriate waveforms \cite{1267994,951379} to mitigate various phenomena such as transmitter synchronization (sync) errors, and intersymbol interference (ISI), among others. The presence of such phenomena affects the performance of the systems at the receiver and consequently can affect the accuracy of computing even in the conventional computing paradigm. Therefore, for OTA computing to be integrated into modern communication systems, its compatibility with digital communication techniques and hardware used in current generations of wireless networks must be facilitated, which is an open research topic to be explored.


\subsection{Literature Review}\label{subsec:Rev}

In recent years, OTA computing has attracted a lot of study due to its ability to perform efficient computations on the data of a large number of devices while utilizing minimal resources. In the seminal works \cite{8550555,6573232,Goldenbaum2015,6125268}, OTA computing was facilitated as a technique to approximate various target functions using their nomographic decomposition. To address the fact that the former decomposition may not be trivial, a deep neural network (DNN) framework for approximating such target functions was proposed in \cite{10506799}. While these works aimed at establishing OTA computing for multivariate function computation, resource allocation studies have also been conducted to optimize the performance of the latter which is critical for any system. Specifically, optimal power policies for OTA computing were studied in \cite{liu_over--air_2020, cao_optimized_2022}, while optimal power allocation schemes and techniques to improve performance under imperfect channel state information (CSI) scenarios were proposed in \cite{OTA_imperfect}.

To further enhance and enable OTA computing, much recent research has focused on its cooperation with other technologies. In this sense, in \cite{Bouzinis2023,jiang_over--air_2019,fang_over--air_2021}, reconfigurable intelligent surfaces (RISs) were proposed to improve performance through optimal handling of the RIS for improved channel conditions and power allocation. In the same direction, the use of unmanned aerial vehicle (UAV) to create optimal channel conditions by solving trajectory optimization problems was investigated in \cite{fu_uav_2021}. In addition, MIMO systems have also been proposed as a way to improve the performance of OTA computing. As such, different directions have been explored in \cite{zhai_hybrid_2021,8468002,MIMO3,9310223,8807380}, including joint hybrid beamforming and zero-force beamforming under a general pool of OTA computing-related constraints associated with mean squared error (MSE) threshold and outage probability. Furthermore, multiple target function computation has been investigated for MIMO systems, where different target functions are computed simultaneously over separate channels with appropriate precoding \cite{MIMO1}. 

In addition to improving OTA computing, some other technologies can be improved by it. One important example is federated learning (FL), which enables distributed training of DNN frameworks by using individual devices to update isolated parts of a DNN and then combining all the updates for a global update at a fusion center (FC). In this context, OTA computing has been studied as an effective way to provide the updates from the distributed devices to the FC while also achieving reduced convergence time \cite{9272666,yang_federated_2020,xu_learning_2021,fan_joint_2022,mohammadi_amiri_machine_2020}. Furthermore, in \cite{nam_conv}, FL convergence was considered when some devices do not participate with updates for each training round, where it was proven that the training will still converge, showcasing the effectiveness of OTA computing when used to enable distributed optimization techniques.

\subsection{Motivation \& Contribution}\label{subsec:Motive}
Although OTA computing has been extensively investigated, either alone or in combination with other technologies, almost all related works are based on the assumption of analog transmission. In \cite{channelcomp}, a framework was proposed to enable OTA computing for digital systems, i.e., to perform the conversion to bits, but the impact of modern communication system components, such as waveform transmission and filters at both ends, was not considered. However, as is known from conventional systems, these components are prone to practical problems such as sync errors and ISI, both of which can degrade system performance. 

In order to facilitate the use of OTA computing in modern devices, it is crucial to investigate the performance of OTA computing under these phenomena, while at the same time aiming to establish its compatibility with modern communication components, which motivates the present work. Furthermore, in relation to the effects of these phenomena, the established waveforms have been proposed for pure communication between devices, where bit error rate (BER) is of interest, meaning that they neglect any differences between the performed tasks, such as computing and conventional data transfer, that are enabled by physical layer communication. Therefore, since computing is of interest, which is more accurately evaluated by MSE, it is reasonable to investigate more appropriate waveforms to mitigate sync errors error and ISI, without assuming that the accuracy of computing is invariant to the implemented waveforms.

In this sense, our work aims to fill the gap regarding the performance of OTA computing when used in modern communication systems that utilize narrowband channels. At the same time, we aim to improve the accuracy of OTA computing by proposing a method to generate waveforms that can better handle phenomena that affect currently deployed systems. Therefore, the contribution of our work is summarized in the following points:

\begin{itemize}
    \item We provide a theoretical analysis of the effect of sync errors and ISI when the transmitted waveforms are incorporated into the system model of OTA computing. For this purpose, we focus on the raised cosine (RC) and better-than-raised-cosine (BTRC) waveforms which are commonly used in modern communication systems. The statistical properties of the established waveforms are also studied, and approximations are proposed to study the behavior of the waveforms in order to study the average MSE of the OTA computing system. 

    \item In order to mitigate the effect of sync errors and ISI on the performance of OTA computing, we provide a thorough analysis, formulating optimization problems to address different conditions. Specifically, sync errors are studied separately and also in combination with ISI when OTA computing is used. In both scenarios, optimal power allocation policies, which are applicable to any waveform, are extracted by checking all critical points of the arising MSE expressions.

    \item 
    For the first time, we propose a novel DNN framework that aims to generate appropriate waveforms to improve the performance of OTA computing. Specifically, a DNN architecture is proposed along with a custom loss function that is introduced to ensure that the DNN-generated waveforms satisfy energy and bandwidth constraints similar to those of the currently implemented waveforms. While seminal works \cite{1267994,951379} related to waveform design focus on simpler models considering only AWGN and ISI and assuming optimal time sampling, our work goes beyond that by incorporating sync errors at the transmitters. To this end, the training of the DNN includes channel fading and noise as well as sync errors and ISI, allowing the DNN to mitigate their combined effects. Moreover, the proposed framework also takes into account the division of the used waveforms between the transmitter and the receiver side, thus correctly encompassing the design structure of modern transceivers.

    \item Based on the theoretical part of our work and the DNN-generated waveforms, simulation results are presented for varying conditions, including increasing number of transmitting devices, transmit signal-to-noise ratio (SNR), and sync errors variance. Simulations are performed for RC, BTRC and the DNN-generated waveform under the extracted optimal power allocation to compare the performance of the currently utilized waveforms. The results showcase that the DNN-generated waveform achieves considerable performance gain over RC and BTRC, emphasizing the significance of the proposed framework. 
\end{itemize}

\subsection{Structure}\label{sec:structure}
The remainder of this paper is organized as follows. Section \ref{sec:sysMod} describes the system model, presenting the basic concepts of OTA computing, the currently used waveforms, and their combination. In Section \ref{sec:theoryOpt}, we formulate the optimization problems of minimizing the average MSE in the presence of sync errors and ISI, and propose corresponding solutions. In Section \ref{sec:DNNWaveforms}, we discuss the proposed DNN framework and how it incorporates energy and bandwidth constraints. In Section 
\ref{sec:Results}, we present simulation results and discussion on the performance of OTA computing and the utilization of the proposed DNN-generated waveforms, while Section \ref{sec:conclusions} concludes the work.

\subsection{Notation}\label{sec:notation}
From now on, vectors are denoted by bold lowercase letters. Sets and sequences are denoted by $\{ \cdot \}$. The expectation of a random multivariate expression w.r.t. the random variable $X$ is denoted by $\mathbb{E}_{X}[\cdot]$, while the discrete Fourier transform (DFT) operation is denoted by $\mathcal{F}(\cdot)$.

\section{System Model} \label{sec:sysMod}

\subsection{OTA Computing Preliminaries}\label{subsec:OTAP}
In our work, we consider an OTA computing system consisting of a receiver, which acts as an FC, and multiple transmitting devices. Let $K$ be the number of transmitting devices in the OTA computing system, while the measurements of all devices are independent.  Assume that we want to calculate a function $f:\mathbb{R}^{K} \rightarrow \mathbb{R}$ of all transmitted data, denoted as $f(x_1, x_2, \cdots, x_K)$. When $f$ is a nomographic function, it is known that there is an appropriate pre-processing function $\varphi_{k}:\mathbb{R} \rightarrow \mathbb{R}, \forall k \in \{1, \cdots, K \}$ and a post-processing function $\psi:\mathbb{R} \rightarrow \mathbb{R}$ such that the target function $f$ is given by
\begin{equation} \label{eq:Nomographic}
f(x_{1,t}, x_{2,t}, \cdots, x_{K,t}) = \psi\left(\sum_{k=1}^{K} \varphi_{k}(x_{k,t})\right),
\end{equation}
where $x_{k,t}$ is the data sample of the $k$-th device at the $t$-th time instance. Due to the stochastic nature of the wireless medium, all transmitted data are subject to channel fading and noise at the receiver, resulting in 
\begin{equation} \label{eq:NonIdealNomographic}
\hat{f} = \psi\left(\sum_{k=1}^{K}h_k \varphi_{k}(x_{k,t}) + n\right),
\end{equation}
where $h_k$ denotes the narrowband block flat-fading channel of the $k$-th device \cite{liu_over--air_2020,fang_over--air_2021}, and $n$ denotes the additive white Gaussian noise (AWGN) with $\mathbb{E}_{n}[n] = 0$ and $\mathbb{E}_{n}[n^2] = \sigma^2$, where $\sigma^2$ is the noise power. We define the set of all devices as $\mathcal{K} = \{ 1, \cdots, K \}$, where the devices are ordered in ascending order of their channel gains. For the transmitted data of each device it is assumed that both $\mathbb{E}_{x_k}[x_{k,t}] = 0, \,\, \forall t$ and $\mathbb{E}_{x_k}[x_{k,t}^2] = 1, \,\, \forall t$ hold. Without loss of generality, we assume that the receiver and all transmitting devices are equipped with a single antenna. We assume that perfect CSI is available at both the transmitter and the receiver.

\subsection{Overview of Basic Waveforms}\label{subsec:waveformBasic}
Modern communication systems rely on the use of appropriate waveforms to deal with the effects of phenomena such as ISI, that are provoked by the limited bandwidth availability and the channels' frequency selectivity. In general, ISI occurs when the currently transmitted symbols are interfered with past and future symbols. As a result, the base station (BS) may not be able to correctly reconstruct the original symbol. Let $T$ be the symbol period of all devices and $z_{k}(t)$ be the waveform associated with the $k$-th device and its data $x_{k,t}$. Because of its ability to mitigate ISI, one of the most commonly implemented waveforms for modern digital systems is the RC waveform, expressed in the time domain as
\begin{equation}\label{eq:RC}
    z_{\mathrm{RC}}(t) = \frac{1}{T}\mathrm{sinc}\left( \frac{t}{T} \right) \frac{\cos\left( \pi \alpha \frac{t}{T} \right)}{1-\left( 2\alpha\frac{t}{T} \right)^2} ,
\end{equation} 
where $ 0 \leq \alpha \leq 1$ is the roll-off factor. Another widely used waveform that is known to provide better performance than the raised cosine in single-user scenarios is the BTRC waveform \cite{1267994,951379}, which is the theoretical performance benchmark of waveforms that satisfy Nyquist's criterion. The BTRC waveform is given in the time domain as
\begin{equation}\label{eq:fExp}
    z_{\mathrm{BTRC}}(t) \hspace{-0.08cm}=\hspace{-0.08cm} \frac{1}{T}\mathrm{sinc}\hspace{-0.05cm}\left(\hspace{-0.05cm} \frac{t}{T} \hspace{-0.05cm}\right) \hspace{-0.07cm}\frac{4\beta \pi t \sin{\left( \frac{\pi \alpha t}{T} \right)} \hspace{-0.05cm}+\hspace{-0.05cm} 2\beta^2 \cos\left(  \frac{\pi \alpha t}{T} \right) \hspace{-0.05cm}-\hspace{-0.05cm} \beta^2}{(2\pi t )^2  \hspace{-0.05cm}+\hspace{-0.05cm} \beta^2},
\end{equation} 
where $\beta = (2T\ln{2}) / \alpha$. The bandwidth of both waveforms, $W_{\mathrm{SC}}$, is dependent on the selected roll-off factor, i.e., 
\begin{equation}\label{eq:band}
    W_{\mathrm{SC}} = (1+\alpha)W,
\end{equation}
where $W$ is the utilized bandwidth.
In general, the larger the roll-off factor, the better the system's ability to mitigate ISI. However, as observed by \eqref{eq:band}, a larger roll-off factor indicates a larger spectrum allocation, creating a critical trade-off between communication performance and resource efficiency.  

\begin{figure}
    \centering
    \includegraphics[width=0.95\columnwidth]{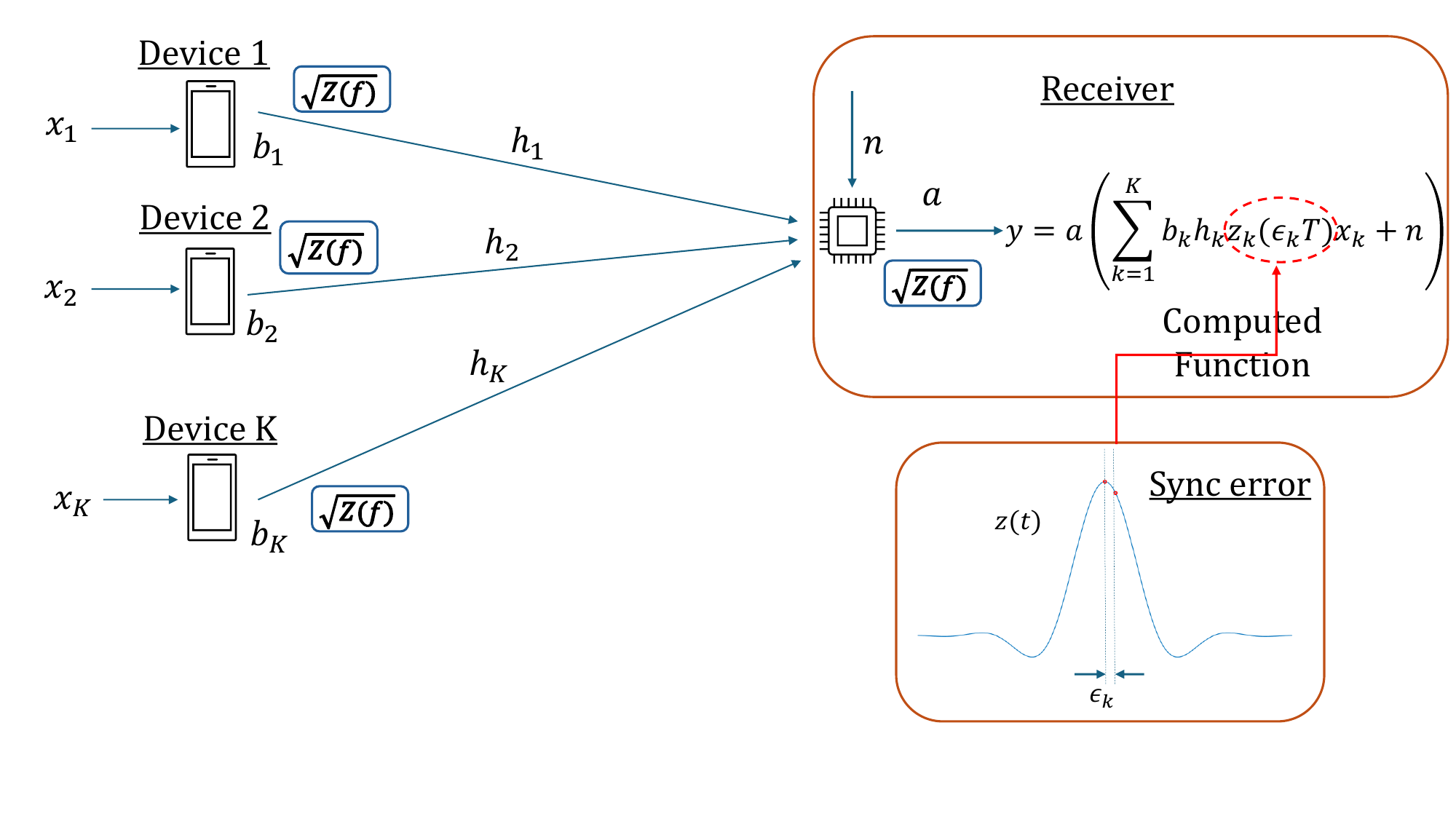}
    \vspace{-6mm}
    \caption{OTA computing model with sync errors.}
    \vspace{-3mm}
    \label{fig:model}
\end{figure}

It should be noted that, although the above waveforms can eliminate ISI, this is only possible with perfect synchronization at the transmitters. Therefore, in practical wireless communication systems, where perfect synchronization is difficult to achieve, using these waveforms does not necessarily result in zero ISI.
Furthermore, it is important to emphasize that modern communication systems mostly rely on a two-part split of the used waveform, i.e., for practical reasons, both the transmitting device and the BS are equipped with the square-root filter of the waveform \cite{10464656} as shown in Fig. \ref{fig:model}, which leads to optimal SNR at the receiver side \cite{1057571}. Note that this technique is the only filter realization that can achieve maximum SNR at the receiver, facilitating it as the optimal filter realization in modern communication systems. 

\subsection{OTA Transmission under Sync Errors and ISI}
Without loss of generality, we assume that the target function is the sum of all transmitted data \cite{liu_over--air_2020,fang_over--air_2021}. In this case, no pre- or post-processing functions are needed, except for appropriate power allocation of the participating devices. Let $b_k \in \mathbb{C}$ be the transmit equalization factor at the $k$-th device such that $|b_k|^2$ denotes the transmit power, $\arg \{b_k\}$ denotes the phase of the transmit signal and the transmitted signal of the $k$-th device in time is given as $x_kb_kz(t)$. Similarly, $a \in \mathbb{C^*}$ denotes the receiver gain factor. All transmitting devices are assumed to have a common maximum power magnitude $P_{\mathrm{max}}$, so that $|b_k|^2E \leq P_{\mathrm{max}}$ for all $k \in \mathcal{K}$, where $E = \int_{-\infty}^{\infty} |z(t)|^2 \mathrm{d} t$
is the waveform energy, which is unitless. This constraint can be equivalently written as $|b_k| \leq \sqrt{P}$, where $P = P_{\mathrm{max}} / E$. Due to the perfect CSI availability, the phase of $b_{k}$ can always be chosen in such a way that the phase shift introduced by the fading is always eliminated. Thus, we consider that the receiver gain, the transmit power, and the channel coefficients are all real numbers, i.e., $a, b_k, h_k \in \mathbb{R},\,\, \forall k \in \mathcal{K}$.


For our analysis, we consider that the received signal is subject to two dominant occurring errors, namely the sync errors, i.e. \textit{symbol timing errors} \cite{951379}, and the existence of ISI. For the sync errors, $\epsilon_k = t / T, \forall k \in \mathcal{K}$, we assume that they follow i.i.d. Gaussian distributions centered around the ideal sync time \cite{1446870} each with variance $\sigma_{\epsilon_k}^2$, thus $\epsilon_k \sim \mathcal{N}(0,\sigma_{\epsilon}^2)$  \footnote{In the case of a common receiver timing offset that follows a Gaussian distribution with zero mean and some variance, each device can be considered to follow a non-i.i.d. Gaussian distribution, resulting from the sum of the two distributions, with a total sync error $\epsilon_k' = t / T, \forall k \in \mathcal{K}$ such that $\epsilon'_k \sim \mathcal{N}(0,\sigma_{\epsilon'}^2)$ occurs and the analysis is the same.}. If the sync errors are the only imperfections at the transmitters, 
the received signal can be described as
\begin{equation}\label{eq:RecNoISI}
    \hat{y} = a \left( \sum_{k=1}^{K}x_{k,0}z_k(\epsilon_k T)b_kh_k + n \right).
\end{equation}
On the other hand, if the sync errors coexist with ISI, the received signal is given as
\begin{equation}\label{eq:RecISI}
    \begin{aligned}
    \hat{y}_{\mathrm{ISI}} &= a \left( \sum_{k=1}^{K}x_{k,0}z_k(\epsilon_k T) b_kh_{k} + n \right) \\
    & \quad + a\underbrace{ \left( \sum_{\substack{q = - \frac{\mu}{2}  \\
    q \neq 0}}^{ \frac{\mu}{2} } \sum_{k=1}^{K}x_{k,q}z_k(T(q + \epsilon_k))b_kh_{k} \right)}_{\text{ISI terms}},
    \end{aligned}
\end{equation}
where $\mu$ is assumed to be even and denotes the number of ISI symbols affecting the currently transmitted symbol, while $h_{k}$ is the narrowband channel coefficient of the $k$-th device for the symbol $x_{k,q}$, which is transmitted at the $q$-th time instance relative to the current time, denoted by $t=0$ for convenience. It is also assumed that $t=0$ corresponds to the ideal sync time. Note that in both cases the ideally received signal is given by
\begin{equation}\label{eq:RecIdeal}
    r = \sum_{k=1}^{K}x_{k,0}. 
\end{equation}

\begin{figure}
    \centering
    \includegraphics[width=0.8\columnwidth]{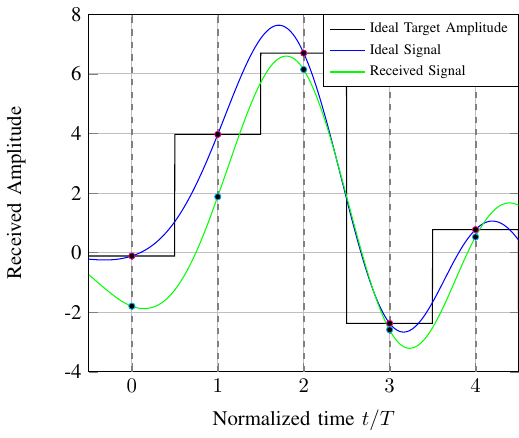}
    \caption{Signal amplitude of the target signal \eqref{eq:RecIdeal} and received signal \eqref{eq:RecISI} with sync errors with $\sigma_{\epsilon} = 0.1$ for RC waveform of $\alpha = 0.3$ in the absence of noise. The samples for each case are depicted as bullet points. For the rest of the selected parameters, see Section \ref{sec:Results}.}
    \label{fig:signals}
\end{figure}

\section{Theoretical Analysis}\label{sec:theoryOpt}
In this section, we theoretically investigate the MSE of OTA computing between the ideal and actual received signal for both the RC and the BTRC waveforms. To this end, both sync errors and ISI are considered, which are the dominant types of errors, while the optimal power allocation for any waveform during the OTA transmission is also extracted. 

\subsection{OTA Computing under Sync Errors}\label{sub:SingleNoISI}
By definition, ideal sampling of any waveform occurs exactly at intervals of symbol period $T$. Since we assume that ideal sampling occurs at $t=0$, it is clear that introducing sync errors $\epsilon_k$ will lead to \eqref{eq:RecNoISI}. Consequently, the MSE in this scenario can be expressed as 
\begin{equation}\label{eq:MSEWithoutISI}
    \mathrm{\overline{MSE}}(a,\bm{b}) = \mathbb{E}_{\epsilon_k}\left[\sum_{k=1}^{K} \big( az_k(\epsilon_k T)b_kh_k - 1 \big)^2 + \sigma^2a^2 \right],
\end{equation}
where the expectation w.r.t. to the noise $n$ and $x_k, \,\forall k \in \mathcal{K}$ has been calculated and $\bm{b} = [b_{1}, \cdots, b_{K}]$ is the transmission power vector of all users. To study the performance of the MSE as described by \eqref{eq:MSEWithoutISI}, we have to first calculate the statistics of the waveform w.r.t. the sync errors. In order to simplify this calculation we consider a geometric series based approximation of the fraction term of the expression for RC in \eqref{eq:RC}, which allows us to write
\begin{equation}\label{eq:approxFrac}
    \frac{1}{1-4 \alpha^2 \epsilon_k^2} = \sum_{n=0}^{\infty} (4 \alpha^2 \epsilon_k^2)^n = 1 + 4 \alpha^2 \epsilon_k^2 + \mathcal{O} (16\alpha^4 \epsilon_k^4),
\end{equation}
where $\mathcal{O}(\cdot)$ denotes the summary of the least-significant terms which can be considered negligible. 
The first equality holds due to the convergence of the geometric series whenever $\epsilon_k < 0.5$ and the approximation is strong due to the small values of the roll-off factor and the sync errors leading to negligible values as the powers of the sync errors $\epsilon_k$ increase. 

Thus, the mean amplitude, $\bar{\epsilon}_{1}$, and the mean squared amplitude, $\bar{\epsilon}_{2}$, at of the desynchronized waveform can be strongly approximated as described by the following proposition.
\begin{proposition}\label{lemma:lemma1}
    The mean amplitude value and the mean squared amplitude value of the desynchronized transmission for device $k\in\mathcal{K}$ and sampled at $t=0$ are strongly approximated by 
\begin{equation}\label{eq:sincE}
    \bar{\epsilon}_{1} 
    = 1 \hspace{-0.1cm}+ \hspace{-0.15cm}\sum_{m=1}^{\infty} \hspace{-0.075cm}\kappa_{m} \sigma_{\epsilon}^{2m}(2m-1)!! \hspace{-0.05cm}+ \hspace{-0.075cm}4\alpha^2 \hspace{-0.075cm}\sum_{m=0}^{\infty} \hspace{-0.075cm}\kappa_{m} \sigma_{\epsilon}^{2m+2}(2m+1)!! ,
\end{equation} 
where $\kappa_m$ denotes the coefficients given as
\begin{equation}\label{eq:coefficientSinc}
    \kappa_m \hspace{-0.6mm} = \sum_{n=0}^{m} \frac{(-1)^{m}\pi^{2m}\alpha^{2(m-n)}}{(2n+1)!(2m-2n)!}
\end{equation}
and 
\begin{align} \label{eq:sincESq}
    \bar{\epsilon}_{2} 
    &= 1 + \sum_{p=1}^{\infty} \lambda_{p} \sigma_{\epsilon}^{2p}(2p-1)!! + \hspace{-0.075cm}8\alpha^2 \hspace{-0.075cm}\sum_{p=0}^{\infty} \hspace{-0.075cm}\lambda_{p} \sigma_{\epsilon}^{2p+2}(2p+1)!! \nonumber \\
    & \qquad + \hspace{-0.075cm}16\alpha^4 \hspace{-0.075cm}\sum_{p=0}^{\infty} \hspace{-0.075cm}\lambda_{p} \sigma_{\epsilon}^{2p+4}(2p+3)!!,
\end{align} 
where $\lambda_p$ denotes the coefficients given as
\begin{equation}\label{eq:coefficientSq}
    \lambda_p \hspace{-0.6mm} = \hspace{-0.6mm} \sum_{n=0}^{p} \sum_{l=0}^{p-n} \sum_{k=0}^{p-n-l} \hspace{-2.3mm}\frac{(-1)^{p}\pi^{2p}\alpha^{2(p-l-n)}}{(2n+1)!(2l+1)!(2k)!(2(p-n-l-k))!} \!.
\end{equation}
\end{proposition}
\begin{IEEEproof}
    The proof is presented in Appendix \ref{app:lemma1}.
\end{IEEEproof}
Then, taking the expected value of \eqref{eq:MSEWithoutISI} w.r.t. the sync errors $\epsilon_k$ occurring in the OTA computing signal, it is straightforward to prove that the MSE under sync errors $\mathrm{MSE}(a,\bm{b}) = \mathbb{E}_{\epsilon_k}\left[ \mathrm{\overline{MSE}}(a,\bm{b}) \right]$ is given as  
\begin{equation}\label{eq:MSEWithoutISIfinal}
    \mathrm{MSE}(a,\bm{b}) = \sum_{k=1}^{K} \left( \left(ab_kh_k \right)^2\bar{\epsilon}_{2} +1 \right) -2\sum_{k=1}^{K} ab_kh_k \bar{\epsilon}_{1} + \sigma^2a^2.
\end{equation}

It is important to emphasize that the following power policy allocation technique is not restricted by the use of the approximations given in \eqref{eq:sincE} and \eqref{eq:sincESq} and can be used by numerically calculating $\bar{\epsilon}_{1} = \mathbb{E}_{\epsilon_k}[z_k(\epsilon_k T)]$ and $\bar{\epsilon}_{2} = \mathbb{E}_{\epsilon_k}[z_k^2(\epsilon_k T)]$. However, the strong approximation in \eqref{eq:MSEWithoutISIfinal} shows interest because it allows a direct calculation of the roll-off factors effect on the MSE in contrast to the numerical approach, meaning that it can also be used to solve optimization problems that would include the roll-off factor as part of the optimization variables, for instance in joint MSE-bandwidth problems

The MSE in \eqref{eq:MSEWithoutISIfinal} can now be minimized by finding the optimal power allocation of the OTA transmission. This problem is formulated as
\begin{equation}\tag{\textbf{P1}}\label{eq:optWithoutISI}
    \begin{array}{ll}
    \mathop{\mathrm{min}}\limits_{a,\bm{b}}
    & \mathrm{MSE}(a,\bm{b})\\
    \,\,\textbf{s.t.} \quad
    & \mathrm{C}_1: b_{k} \leq \sqrt{P}, \ \forall k \in \mathcal{K}.
    \end{array}
\end{equation}
The Lagrangian can be written as
\begin{equation}\label{eq:LagrangeNoISI}
\mathcal{L}=\mathrm{MSE}(a,\bm{b})+\sum_{k=1}^K\lambda_k(b_k-\sqrt{P}),
\end{equation}
where $\lambda_k$ are the Lagrange multipliers.

By using the Karush-Kuhn-Tucker (KKT) conditions, the optimal solution must satisfy the following:
\begin{equation}\label{eq:KKTaNoISI}
\frac{\partial \mathcal{L}}{\partial a}=0,\frac{\partial \mathcal{L}}{\partial b_k}=0, \forall k \in \mathcal {K}
\end{equation}
and 
\begin{equation}\label{eq:lagrangeKKTNoISI}
\lambda_k(b_k-\sqrt{P})=0, \forall k \in \mathcal {K},
\end{equation}
which equivalently leads to 
\begin{equation}\label{eq:KKTbNoISI}
\frac{\partial \mathrm{MSE}(a,\bm{b})}{\partial b_k}=0\Rightarrow b_{k} = \bar{\epsilon}_{1} / (ah_k \bar{\epsilon}_{2})\text{ } \mathrm{or} \text{ } b_k=\sqrt{P}, \,\, \forall k \in \mathcal{K}.
\end{equation}
Also, since $\mathrm{MSE}(a,\bm{b})$ is a convex function with respect of $b_k$, it holds that
\begin{equation}\label{eq:powerNoISI}
    b_{k} = \min \left\{ \sqrt{P}, \frac{\bar{\epsilon}_{1}}{ah_k \bar{\epsilon}_{2}} \right\}, \forall k \in \mathcal{K}.
\end{equation}
Based on \eqref{eq:powerNoISI}, to further reduce the size of the set of the potential optimal points, it is sufficient to consider $K$ different cases for the values of $\lambda_k$. More specifically, it is noted that  if there exists a device $(i+1) \in \mathcal{K}$ such that $b_{i+1} = \bar{\epsilon}_{1} / (ah_{i+1} \bar{\epsilon}_{2})$, any device $j \in \{ \mathcal{K} | j > i \}$ can select the same inverse channel-like transmit power due to the ascending channel order. Thus, it holds that if $\lambda_{i+1}= 0$, then $\lambda_j= 0, \forall j>i$. Regarding the $i$-th case, by using \eqref{eq:powerNoISI}, the Lagrangian simplifies to the MSE for the corresponding values of $\lambda_k$ and $b_k$, which is given by
\begin{align}\label{eq:MSEaFind}
   \mathcal{L}_i= \mathrm{MSE}_{i}(a) &= a^2 \left( \sum_{k=1}^{i} Ph^2_k \bar{\epsilon}_{2} + \sigma^2 \right) - 2a\sum_{k=1}^{i} \sqrt{P}h_k \bar{\epsilon}_{1} \nonumber \\
    & \quad + \sum_{k=i+1}^{K} \left(1 - \frac{\bar{\epsilon}_{1}^2}{\bar{\epsilon}_{2}} \right) + i.
\end{align}
It should be noted that for the primal feasibility conditions of \eqref{eq:optWithoutISI}, i.e., $\mathrm{C}_1$, to be satisfied, since the channel gains have been ordered, the following must hold for the receiver gain
\begin{equation}\label{eq:aCond}
    a \geq \frac{\bar{\epsilon}_{1}}{\sqrt{P}h_{i+1} \bar{\epsilon}_{2}}.
\end{equation}
If \eqref{eq:aCond} is satisfied, the minimum value of \eqref{eq:MSEaFind} is reached when $\partial{\mathcal{L}_i}/\partial a=0$, i.e., 
\begin{equation}\label{eq:aMaxINoSI}
    a = a_{i} =  \frac{\sqrt{P}\bar{\epsilon}_{1}\sum_{k=1}^{i} h_k}{P\bar{\epsilon}_{2}\sum_{k=1}^{i} h^2_k + \sigma^2}.
\end{equation}
Then, we can define the set of optimal solutions of $a$ as follows
\begin{equation}\label{eq:setOptimalA}
    \mathcal{A} = \left \{ a = a_{i} \bigg| a_{i}  \geq \frac{\bar{\epsilon}_{1}}{\sqrt{P}h_{i+1} \bar{\epsilon}_{2}}, \forall i \in \mathcal{K} \right \}.
\end{equation}
Observe that if $a \not \in \mathcal{A}$, it cannot be optimal since the KKT conditions described by \eqref{eq:KKTaNoISI} are not satisfied.
Comparing the values of the sequence $\mathrm{MSE}_i(a_{i}), \ \forall i \in \mathcal{K}, \ \forall a_{i} \in \mathcal{A}$, described by \eqref{eq:MSEaFind}, we can identify the number of devices $i^*$ that must transmit with maximum power and is equal to
\begin{equation}\label{eq:criticalNoISI}
    i^* = \underset{\substack{ 1 \leq i \leq K \\ a_{i} \in \mathcal{A}}}{\mathrm{argmin}}\,\, \{ \mathrm{MSE}_{i}(a_{i}) \}.
\end{equation}
Then, the optimal power allocation at the devices and the receiver can be calculated by combining \eqref{eq:criticalNoISI}, \eqref{eq:aMaxINoSI} and \eqref{eq:powerNoISI} in this specific order. 

\subsection{OTA Computing under Sync Errors and ISI}\label{sub:SingleISI}
In limited bandwidth systems, except for sync errors, ISI is also present. Although waveforms such as RC and BTRC have relatively small amplitudes around their roots at multiples of $T$, the effect of ISI cannot be neglected, especially in multiple access schemes such as OTA computing where a large number of connected devices are present in the system. As a result, the study of ISI is important for understanding the performance of OTA computing in practice and for extracting optimal power allocation policies to mitigate its effect. 

In this case the received signal is affected by the currently transmitted signal as well as earlier and later transmissions at time instances $t_k = T(q + \epsilon_k), \forall k \in \mathcal{K}$ where $q \in \{ -  \mu / 2 , \cdots,  \mu / 2  \}$ and the received signal is given by \eqref{eq:RecISI}. Therefore, the MSE now contains additional terms due to ISI terms and is equal to
\begin{equation}\label{eq:recFullISI}
    \begin{aligned}
    \mathrm{\overline{MSE}}^{\mathrm{ISI}}(a,\bm{b}) &= \sum_{k=1}^{K} \left( az_k(\epsilon_k T)b_kh_k - 1 \right)^2 + \sigma^2a^2 \\
    & \quad + \sum_{\substack{q = -  \frac{\mu}{2}  \\
    q \neq 0}}^{ \frac{\mu}{2} } \sum_{k=1}^{K} \left( az_k(T(q+\epsilon_k))b_kh_{k} \right)^2, 
    \end{aligned}
\end{equation}
where $h_{k}$ denotes the channel of the $k$-th device at the $q$-th time instance. Since the samples at different times are independent of each other, $\mathbb{E}_{x_k}[x_{k,q_{1}}x_{k,q_{2}}] = 0$, whenever $q_{1} \neq q_{2}$.
Then, taking the expectation w.r.t. $\epsilon_k$ for the MSE, $\mathrm{MSE}^{\mathrm{ISI}}(a,\bm{b}) = \mathbb{E}_{\epsilon_k}\left[\mathrm{\overline{MSE}}^{\mathrm{ISI}}(a,\bm{b}) \right]$, we derive an expression similar to the one in \eqref{eq:optWithoutISI} described by
\begin{equation}\label{eq:MSE_ISI_q}
    \begin{aligned}
    \mathrm{MSE}^{\mathrm{ISI}}(a,\bm{b}) &\!=\! \sum_{k=1}^{K} \left( \left(ab_kh_k \right)^2\tilde{\epsilon}_{0} +1 \right) - 2\sum_{k=1}^{K} ab_kh_k \check{\epsilon} \\
    & \quad + \sum_{\substack{q = -  \frac{\mu}{2}  \\
    q \neq 0}}^{ \frac{\mu}{2} } \sum_{k=1}^{K} \left(ab_kh_{k}\right)^2\tilde{\epsilon}_{q}  + \sigma^2a^2 ,
    \end{aligned}
\end{equation}
where $\tilde{\epsilon}_{q} = \mathbb{E}_{\epsilon_k}[z_k^2(T(q+\epsilon_k))]$ and $\check{\epsilon} = \mathbb{E}_{\epsilon_k}[z_k(\epsilon_k T)]$. In particular, for the strongest symbol interference caused by the previous and the next symbol of the currently sampled symbol, the mean values can be rigorously calculated in closed form as described by the following proposition.
\begin{proposition}\label{lemma:lemma2}
    The mean squared amplitude of the raised cosine waveform at the time instances corresponding to $q = \pm 1$ with roll-off factor $\alpha=0$ is equal to  
    \begin{equation}\label{eq:alpha0}
        \tilde{\epsilon}_{1} = 1 +\sum\limits_{m=1}^{\infty} \sum\limits_{n=0}^{2m} A_{n}A_{2m-n} +  B_n B_{2m-n} \sigma_{\epsilon}^{2m}(2m-1)!!,
    \end{equation} 
    and $\tilde{\epsilon}_{-1} = \tilde{\epsilon}_{1}$, where the coefficients $A_n, B_n$ are given as
    \begin{equation}\label{eq:coefficientsInitial}
    \begin{split}
      A_n &= \hspace{-1mm}\sum\limits_{l=\frac{n+1}{2}}^{\infty} \hspace{-1mm}\frac{(-1)^l\pi ^{2l}}{(2l+1)!}\binom{2l}{n} \text{ if } n \text{ odd, } A_n = 0 \text{ if } n \text{ even } \\
    B_n &= \sum\limits_{l=\frac{n}{2}}^{\infty} \frac{(-1)^l\pi ^{2l}}{(2l+1)!}\binom{2l}{n} \text{ if } n \text{ even, } B_n = 0 \text{ if } n \text{ odd} .
    \end{split}
    \end{equation}
\end{proposition}
\begin{IEEEproof}
    The proof is presented in Appendix \ref{app:lemma2}.
\end{IEEEproof}


Considering ISI, the new optimization problem to solve can be formulated as
\begin{equation}\tag{\textbf{P2}}\label{eq:optWithISI}
    \begin{array}{ll}
    \mathop{\mathrm{min}}\limits_{a,\bm{b}}
    &\mathrm{MSE}^{\mathrm{ISI}} (a, \bm{b})\\
    \,\,\textbf{s.t.} \quad
    & \mathrm{C}_1: b_{k} \leq \sqrt{P}, \forall k \in \mathcal{K}.
    \end{array}
\end{equation}
The Lagrangian of the latter can be written as
\begin{equation}\label{eq:LagrangeISI}
\mathcal{L}^{\mathrm{ISI}}=\mathrm{MSE}^{\mathrm{ISI}}(a,\bm{b})+\sum_{k=1}^K\lambda_k^{\mathrm{ISI}}(b_k-\sqrt{P})
\end{equation}
where $\lambda^{\mathrm{ISI}}_k$ are the Lagrange multipliers.

By using the KKT conditions, the optimal solution must satisfy the following conditions:
\begin{equation}\label{eq:KKTaISI}
\frac{\partial \mathcal{L}^{\mathrm{ISI}}}{\partial a}=0,\frac{\partial \mathcal{L}^{\mathrm{ISI}}}{\partial b_k}=0, \forall k \in \mathcal {K}
\end{equation}
and 
\begin{equation}\label{eq:lagrangeKKTISI}
\lambda^{\mathrm{ISI}}_k\mathcal{L}^{\mathrm{ISI}}(b_k-\sqrt{P})=0, \forall k \in \mathcal {K}.
\end{equation}
which equivalently leads to 
\begin{equation}\label{eq:KKTbISI}
\frac{\partial \mathrm{MSE}^{\mathrm{ISI}}(a,\bm{b})}{\partial b_k}=0\Rightarrow b_{k} = \check{\epsilon} / (ah_{k} \hat{\epsilon}) \text{ } \mathrm{or} \text{ } b_k=\sqrt{P}, \, \forall k \in \mathcal{K},
\end{equation}
where $\hat{\epsilon} = \sum_{q=-\mu/2}^{\mu/2} \tilde{\epsilon}_{q}$. Also, since $\mathrm{MSE}(a,\bm{b})$ is a convex function with respect of $b_k$, it holds that
\begin{equation}\label{eq:powerISI}
    b_{k} = \min \left\{ \sqrt{P}, \frac{\check{\epsilon}}{ah_{k} \hat{\epsilon} 
  } \right\}, \forall k \in \mathcal{K}.
\end{equation}

For \eqref{eq:optWithISI}, similar to \eqref{eq:optWithoutISI}, based on \eqref{eq:powerNoISI}, to further reduce the size of the set of the potential optimal points, it is sufficient to consider $K$ different cases for the values of $\lambda^{\mathrm{ISI}}_k$. More specifically, it is noted that  if there exists a device $(i+1) \in \mathcal{K}$ such that $b_{i+1} = \check{\epsilon} / (ah_{i+1} \hat{\epsilon})$, any device $j \in \{ \mathcal{K} | j > i \}$ can select the same inverse channel-like transmit power due to the ascending channel order. Thus, it holds that if $\lambda^{\mathrm{ISI}}_{i+1}= 0$, then $\lambda^{\mathrm{ISI}}_j= 0, \forall j>i$. Regarding the $i$-th case, by using \eqref{eq:powerISI}, the Lagrangian simplifies to the MSE for the corresponding values of $\lambda^{\mathrm{ISI}}_k$ and $b_k$, which can be written as
\begin{align}\label{eq:MSEaFindISI}\mathcal{L}^{\mathrm{ISI}}_{i} = \mathrm{MSE}_{i}^{\mathrm{ISI}}(a) &= a^2 \left( \sum_{k=1}^{i} Ph^2_k \hat{\epsilon} + \sigma^2 \right) -2a\sum_{k=1}^{i} \sqrt{P}h_k \check{\epsilon} \nonumber \\
& \quad + \sum_{k=i+1}^{K} \left(1 - \frac{\check{\epsilon}^2}{\hat{\epsilon}} \right) + i.
\end{align}
It should be noted that for the primal feasibility conditions of \eqref{eq:optWithISI}, i.e., $\mathrm{C}_1$, to be satisfied, since the channel gains have been ordered, the following must hold for the receiver gain factor
\begin{equation}\label{eq:aCondISI}
    \frac{\check{\epsilon}}{\sqrt{P}h_{i+1}\hat{\epsilon}} \leq a.
\end{equation}
If \eqref{eq:aCondISI} is satisfied, the minimum value of \eqref{eq:MSEaFindISI} is reached when $\partial{\mathcal{L}^{\mathrm{ISI}}_i}/\partial a=0$, i.e.,
\begin{equation}\label{eq:aMaxISI}
    a = a_{i} = \frac{\sqrt{P}\check{\epsilon}\sum_{k=1}^{i} h_k}{P\hat{\epsilon}\sum_{k=1}^{i} h^2_k + \sigma^2}.
\end{equation} Then, similar to the non-ISI case, we can define the set of optimal solutions of $a$ as follows
\begin{equation}\label{eq:setOptimalA_ISI}
    \mathcal{A}^{\mathrm{ISI}} = \left \{ a = a_{i} \bigg| a_{i}  \geq \frac{\check{\epsilon}}{\sqrt{P}h_{i+1}\hat{\epsilon}} , \forall i \in \mathcal{K} \right \}.
\end{equation}
Observe that if $a \not \in \mathcal{A}^{\mathrm{ISI}}$, it cannot be optimal since the KKT conditions described by \eqref{eq:KKTaISI} are not satisfied.
Comparing the values of the sequence $\mathrm{MSE}^{\mathrm{ISI}}_i(a_{i}), \ \forall i \in \mathcal{K}, \ \forall a_{i} \in \mathcal{A}^{\mathrm{ISI}}$, described by \eqref{eq:MSEaFindISI}, we can identify the number of devices $i^*$ that must transmit with maximum power and is equal to
\begin{equation}\label{eq:criticalISI}
    i^* = \underset{\substack{ 1 \leq i \leq K \\ a_{i} \in \mathcal{A}^{\mathrm{ISI}}}}{\mathrm{argmin}}\,\, \{ \mathrm{MSE}^{\mathrm{ISI}}_{i}(a_{i}) \}.
\end{equation}
Then, the optimal power allocation at the devices and the receiver can be calculated by combining \eqref{eq:criticalISI}, \eqref{eq:aMaxISI} and \eqref{eq:powerISI} in this specific order. 


It is clear from \eqref{eq:powerISI} and \eqref{eq:aMaxISI} that the MSE of an OTA transmission is affected by the statistics of the selected waveform. Thus, it is worth emphasizing that both power policies discussed in this section hold for any waveform because they depend on the statistical moments of a waveform rather than their explicit closed-form expression, which makes them general. This also allows them to be used for MSE minimization with any waveform, including RC and BTRC, and to perform near-optimal waveform design for OTA transmission via a custom DNN implementation.

\section{DNN-generated Waveforms}\label{sec:DNNWaveforms}
In this section, we utilize the optimal transmission policy derived in Section \ref{sub:SingleISI} to find a waveform tailored to the goal of OTA computing. Although in theory a waveform can expand infinitely in time, in practice, a windowed version of a waveform is transmitted, meaning that waveforms only extend to a limited time window. The larger the time window, the more ISI symbols $\mu$ must be considered. Therefore, we are interested in finding a waveform that minimizes the MSE given in \eqref{eq:recFullISI}, while simultaneously considering practical conditions and constraints during design. 

\subsection{Waveform Design Problem Formulation}\label{subsec:waveforProb}
In addition to minimizing MSE, which is the primary goal of OTA computing, waveform design must consider associated constraints, such as energy and bandwidth. Based on these and denoting by $\bm{h} = [h_1, h_2, \cdots, h_{K}]$ the channels of the devices and by $\bm{x}_0 = [x_{1,0}, x_{2,0}, \cdots, x_{K,0}]$ the transmitted data at the current transmission time, the following problem is formulated
\begin{equation}\tag{\textbf{P3}}\label{eq:optMSEWave}
    \begin{array}{ll}
    \mathop{\mathrm{min}}\limits_{z(t)}
    & \!\!\mathbb{E}_{\epsilon_k,\bm{h}, \bm{x}_0,n} \left[\left|\hat{y}_{\mathrm{ISI}} - r\right|^2 \right]\\
    \text{s.t.}& \!\! \mathrm{C}_1: \int_{-\infty}^{\infty} |Z(f)| df = E, \\
     & \!\! \mathrm{C}_2 :|Z(f)| = 0, \,\, |f| \geq (1+\alpha)W , \\
    & \!\! \mathrm{C}_3: z(t) = z(-t), \,\,0 \leq t \leq \frac{(\mu+1)T}{2},
    \end{array}
\end{equation}
where $Z(f)$ denotes the continuous Fourier transform of the waveform $z(t)$ that must be found. Constraint $\mathrm{C}_1$ is written in the frequency domain leveraging Plancherel's theorem and relates to the waveform's energy, because each transmitter and the receiver utilize the square root of the waveform, $\sqrt{|Z(f)|}$ matched filter, so that the overall response is given as $|Z(f)|$, while $\mathrm{C}_2$ concerns bandwidth and $\mathrm{C}_3$ ensures symmetry in time. 

Examining \eqref{eq:optMSEWave}, it can be observed that $\mathrm{C}_1$ and $\mathrm{C}_2$ are non-linear, while the objective function is highly non-convex due to the expectation taken w.r.t. the sync errors $\epsilon_k$ and channel fading $h_k$ of each device, thus a closed-form expression to find an optimal continuous waveform is intractable. With this in mind, studying a discretized windowed version of the waveform is proposed to better handle the problem. Assuming that $N_s$ time instances are used to discretize $\mu+1$ time periods of the generated waveform, the waveform expands in the time domain from $- (\mu+1)T / 2  $ to $ (\mu+1)T / 2$. Then, the time resolution of the waveform $z$ is $\Delta t = (\mu+1) T / (N_s-1)$ and the time instances correspond to $t = m \Delta t$, with $-(N_s - 1) / 2 \leq m \leq (N_s - 1) / 2 $ and $m \in \mathbb Z$. For practicality, we assume that no sync error greater than half a period can occur, i.e., the desynchronization time instances are limited in the interval $-T / 2 \leq t \leq T / 2$ or equivalently $-(N_s - 1) / (2(\mu + 1)) \leq m \leq (N_s - 1) / (2(\mu + 1))$. 

To discretize the objective function of \eqref{eq:optMSEWave}, equivalent probability mass functions (PMFs) for each sync error can be used to approximate each sync error distribution. To simplify the expression of the objective function, the use of a single receiver sync error $\epsilon$ with PMF $f_{\epsilon}[m]$, instead of multiple errors each with its own PMF, can be achieved by means of the following Proposition \ref{prop:equiv}. 

\begin{proposition} \label{prop:equiv}
    Considering a receiver error $\epsilon$ instead of multiple transmitter errors $\epsilon_k$ yields the same MSE given by \eqref{eq:MSE_ISI_q}.
\end{proposition}
\begin{IEEEproof}
    The proof is presented in Appendix \ref{app:lemma3}
\end{IEEEproof}

Then, leveraging the discrete counterpart of Fourier transform, i.e., fast Fourier transform (FFT), \eqref{eq:optMSEWave} can be rewritten as
\begin{equation}\tag{\textbf{P4}}\label{eq:optMSE_reform}
\begin{array}{ll}
    \mathop{\mathrm{min}}\limits_{\bm{z}}
    & \!\! \mathbb{E}_{\epsilon, \bm{h},\bm{x}_0,n} \left[\left| y[m] - r \right|^2\right] \\
    \text{s.t.}& \!\!  \mathrm{C}_1: \sum_{n=0}^{N-1} |\mathcal{F}( z[m])[n]| = E, \\
    & \!\! \mathrm{C}_2 :|\mathcal{F}( z[m])[n]| = 0,  \, \hspace{-0.25mm} N_{t}(\alpha)  <  n  < N, \\
    & \!\! \mathrm{C}_3: z[m] = z[-m], \, 0 < m \leq \frac{N_s - 1}{2},
\end{array}
\end{equation}
where $\mathcal{F}(\cdot)$ denotes the DFT and the received signal from the generated waveform at the $m$-th time instance is given, similarly to \eqref{eq:RecISI}, as
\begin{equation}\label{eq:discreteSignal}
    \begin{aligned}
    y[m] &= a \left( \sum_{k=1}^{K}x_{k,0}z_{k}[m]b_kh_{k} + n \right) \\
    &+ a\underbrace{ \left( \sum_{\substack{q = -  \frac{\mu}{2}  \\ q \neq 0}}^{ \frac{\mu}{2} } \sum_{k=1}^{K}x_{k,q}z_{k}[m+qT]b_kh_{k} \right)}_{\text{ISI terms}}.
    \end{aligned}
\end{equation}
As observed, the constraints of \eqref{eq:optMSEWave} correspond with those of \eqref{eq:optMSE_reform}, thus making the discrete form of the waveform design problem equivalent to its continuous one.
Regarding constraint $\mathrm{C}_2$ in \eqref{eq:optMSE_reform}, $N_t(\alpha)$ denotes the FFT sample after which this constraint must be enforced, while $N = N_s+N_s^{\mathrm{ZP}}$ denotes the total number of samples used for the FFT, where $N_s^{\mathrm{ZP}}$ denotes the number of additional samples that can be used to capture a zero-padding effect and can be used to obtain a greater frequency resolution in the waveform frequency response. Note that the constraints of \eqref{eq:optMSEWave} correspond with those of \eqref{eq:optMSE_reform}.

Although \eqref{eq:optMSE_reform} contains a finite number of variables, which is easier to tackle via conventional optimization techniques, the expectation in the objective function cannot be expressed in closed form and must be evaluated through a Monte-Carlo sampling approach, since it depends on the joint distribution of the sync errors $\epsilon$, channels $\bm{h}$ and transmitted data $\bm{x}_0$. 
Moreover, due to the terms $|\mathcal{F}( z[m])[n]|$, the energy and frequency constraints are non-linear and non-convex, which also increases the complexity of the formulated problem. With these in mind, we propose the use of a DNN architecture, which can approximate any function \cite{HORNIK1989359}, to design an optimal waveform. This waveform is generated once, prior to deployment, and does not require the involvement of the DNN during the real-time operation of the system. 

\subsection{DNN Framework Approach}
Let $\bm{w}$ denote the weights of the DNN and $z_{\bm{w}}$ denote the output of the DNN model, where each element of the output vector $z_{\bm{w}}$ corresponds to a specific time instance of the desired discretized windowed waveform. To express the objective of \eqref{eq:optMSE_reform} during training, supervised learning can be utilized by feeding vector $\bm{u}= [a, \bm{b}, \bm{h}, \bm{x}_0] \in \mathbb{R}^{3K+1}$ as input to the DNN. Note that the use of transmitted data as input allows the framework to capture the statistics of transmitted data without limiting its range when used in practice. At the same time, it allows the MSE to be accurately described by the received signal and the target signal, both of which require some transmitted data. It should be noted that we initialize the power allocation $a,\bm{b}$ with the values that correspond to the same OTA transmission without sync errors and ISI as in \cite{liu_over--air_2020}. The rationale behind this choice is that, as shown in Section \ref{sec:theoryOpt}, finding the optimal $a,\bm{b}$ requires the statistics of the waveform, which are unknown a priori. Nevertheless, we note that to test the extracted waveform, after the DNN has converged, we obtain the optimal power allocation for the waveform through the analysis of Section \ref{sec:theoryOpt}, specifically \eqref{eq:powerISI}, \eqref{eq:aMaxISI}, \eqref{eq:criticalISI}, and then calculate its MSE. 

 
The architecture of the designed DNN consists of three fully connected hidden layers, each consisting of multiple neurons and implementing the rectified linear unit (ReLU) activation function.  To train the DNN, a training dataset is generated, denoted as $\{\bm{u}_d, r_d \}_{d \in \mathcal{D}}$, where $\mathcal{D}$ is the set of all training samples, $\bm{u}_d$ is the $d$-th input sample and $r_d = \sum_{k=1}^{K} x_{k,0}^{(d)}$ is the target value associated with the $\bm{x}_0$ component of the $d$-th sample. Then, the received signal from the generated waveform at the $m$-th time instance is accurately described, similarly to \eqref{eq:RecISI}, as
\begin{equation}\label{eq:dnnSignal}
    \begin{aligned}
    y_{\mathrm{DNN}}[& \bm{u}_d,m] = a^{(d)} \left( \sum_{k=1}^{K}x_{k,0}^{(d)}z_{k_{\bm{w}}}[m]b_k^{(d)}h_{k}^{(d)} + n \right) \\
    &+ a^{(d)}\underbrace{ \left( \sum_{\substack{q = -  \frac{\mu}{2}  \\ q \neq 0}}^{ \frac{\mu}{2} } \sum_{k=1}^{K}x_{k,q}^{(d)}z_{k_{\bm{w}}}[m+qT]b_k^{(d)}h_{k}^{(d)} \right)}_{\text{ISI terms}},
    \end{aligned}
\end{equation}
where $a^{(d)}, b_k^{(d)}, h_k^{(d)}$ are given by the associated $a, \bm{b}, \bm{h}$ components of the $d$-th sample, $\bm{u}_d$. It should be noted that since the input data at different time instances are unknown, $x_{k,q}$ are passed as arguments to the loss function, but cannot be considered as inputs to the DNN framework itself. However, since both take values from known distributions, the training phase allows the DNN to capture their general statistics.

For the DNN to generate a valid and feasible waveform, the DNN-based waveform must meet the same criteria as the waveforms studied in the literature, as shown in \eqref{eq:optMSE_reform}. Therefore, it is crucial to ensure that the DNN-based waveform has the same energy and bandwidth as the other waveforms to ensure fairness in the utilization of resources in terms of energy and frequency. Both of these constraints must be included in the loss function of the DNN model. Regarding the MSE of the OTA transmission, a challenge arises due to the stochastic nature of the sync errors. In particular, since the sync errors are random, it is not known a priori which value of the waveform should be used in \eqref{eq:dnnSignal}. In essence, to model the time synch error and considering the time discretization of the DNN's output, a random index of the DNN's output should be chosen during each forward pass and the waveform value corresponding to this index should be used in \eqref{eq:dnnSignal}. However, this approach causes the gradient of the output to be lost during the forward propagation and the DNN cannot be trained. To overcome this problem, the  PMF $f_{\epsilon}[m]$ of the sync error distribution for all time instances considered by the DNN's output is used, which allows for a precise calculation of the MSE without gradient loss.  Specifically, based on the assumption that no sync error greater than half a period can occur, $f_{P}[m]$ is considered non-zero only for $-(N_s - 1) / (2(\mu + 1)) \leq m \leq (N_s - 1) / (2(\mu + 1))$.  Then, the loss function of the DNN can be given by
\begin{equation}\label{eq:lossMSE}
    \mathcal{L}_{\mathrm{MSE}} = \frac{1}{|\mathcal{D}|}\sum_{d \in \mathcal{D}} \sum_{m = - \frac{N_s - 1}{2(\mu + 1)}}^{\frac{N_s - 1}{2(\mu + 1)}} \!\! f_{\epsilon}[m] \big(y_{\mathrm{DNN}}[\bm{u}_d,m] \!-\! r_d \big)^2.
\end{equation}

\begin{remark}
Using Proposition \ref{prop:equiv}, \eqref{eq:lossMSE} is subject to only one PMF of $(N_s - 1) / (\mu + 1) + 1$ samples. On the contrary, the expression of the multiple transmitting sync errors described by a multivariate PMF of independent univariate PMFs each with $(N_s - 1) / (\mu + 1) + 1$ would lead to ${\left[(N_s - 1) / (\mu + 1) + 1\right]}^{K}$ possible combinations which is computationally inefficient. Thus, utilizing Proposition \ref{prop:equiv} also lowers the complexity of the calculation of the MSE loss function.
\end{remark}

The utilized bandwidth, as expressed by the discrete Fourier transform (DFT) of the waveform, which gives a sampled form of the continuous frequency response, must first be considered for the bandwidth constraint $\mathrm{C}_2$ in \eqref{eq:optMSE_reform}. Considering the time resolution $\Delta t$, the frequency resolution can be found equal to $\Delta f = 1 / ((\mu+1) T) - 1 / ((\mu+1) T N_s) \approx 1 / ((\mu+1) T)$, since the number of output nodes as well as the overall studied time duration are in general large. 
As is well known, for a baseband signal such as the waveforms described by \eqref{eq:RC} and \eqref{eq:fExp}, the utilized bandwidth is given as $W = 1 / (2T)$. 

Thus, the effect of the roll-off factor on the total bandwidth spread can be studied for a resolution step of $1 / ((\mu+1) T)$, with $\alpha = 0$ corresponding to the $\mu / 2 $-th sample, since then we obtain bandwidth utilization $W = 1 / (2T)$, and $\alpha = 1$ corresponding to the $\mu$-th sample of the DFT, since the utilized bandwidth is $W = 1 / T$. Obviously, this has a restrictive effect on the roll-off factors that can be investigated, thus a way to increase the frequency resolution is needed. This is achieved by zero-padding, which expands a signal in time to achieve the desired frequency resolution, allowing an accurate representation of the frequency response. Denoting the additional duration of the waveforms due to zero-padding as $T^{\mathrm{ZP}}$, the frequency resolution is increased accordingly in $\Delta f^{\mathrm{ZP}} \approx 1 / ((\mu+1) T +  T^{\mathrm{ZP}})$ which can be derived as before but using the extended time duration of the signal. In this way, the number of roll-off factors to be examined can be chosen so that the frequency resolution corresponds to the desired roll-off factor step. For example, if the examined roll-off factors have a step of $\Delta \alpha = 0.1$ corresponding to a desired frequency resolution equal to $\Delta W = \Delta \alpha / (2T)$, we can choose $ T^{\mathrm{ZP}}$ to satisfy 
\begin{equation}\label{eq:DFTcond}
    \frac{1}{(\mu+1) T +  T^{\mathrm{ZP}}} \!=\! \frac{\Delta \alpha}{2T} \Leftrightarrow  T^{\mathrm{ZP}} \!=\! \left( \frac{2}{\Delta \alpha}-\mu-1 \right)T,
\end{equation}
and since the symbol period $T$ corresponds to a number of $(N_s-1) / (\mu+1)$ samples, it is straightforward to calculate the desired number of samples needed to perform zero-padding which are denoted as $N_s^{\mathrm{ZP}}$ and for which $T^{\mathrm{ZP}} = N_s^{\mathrm{ZP}} \Delta t$ holds. Therefore, to ensure that the generated waveform has the same bandwidth utilization as the other waveforms, a constraint is applied to keep the frequency response outside the desired spectrum as close to zero as possible. Note that if the term of the parentheses in \eqref{eq:DFTcond} is negative then the duration of the signal is enough to capture the desired frequency resolution and no zero-padding is needed, resulting in $N_s^{\mathrm{ZP}} = 0$.

\begin{algorithm}[t]\label{alg:WaveformDes}
\caption{DNN Training and Waveform Design}
\textit{Training Phase:} \\

\textbf{Input:} Dataset $\mathcal{D}$ breaks into batches $\mathcal{B}$ with input samples $\bm{u}_d, d \in \mathcal{B}$. \\

Fix number of epochs, $epoch$. \\

\For{$j = 1:\textit{epoch}$}{
Find weights $\bm{w}$ such that
$\mathop{\mathrm{min}}\limits_{\bm{w}}{\mathbb{E}[\mathcal{L}_{tot}(z_{\bm{w}})]}, \forall \bm{u}_d \in \mathcal{B}$. \\

Repeat for all batches. \\
}

\textbf{Output:} Trained DNN with near-optimal weights $\bm{w}^*$. \\

\textit{Waveform Generation for Real-time Use:} \\
Find $z_{\bm{w}^*}$ for every batch $\mathcal{B}$.
Average over all generated $z_{\bm{w}^*}$ to get final waveform $z$. \\
Fit generated waveform to get a real-time applicable waveform, as described in \eqref{eq:generalFit}.
\end{algorithm}
With these in mind, we use the following as part of the loss function
\begin{equation}\label{eq:DFTloss}
    \mathcal{L}_{f} = \frac{1}{|\mathcal{D}|} \sum_{d \in \mathcal{D}} \sum_{n > N_{t}(\alpha)}^{N_s +  N_s^{\mathrm{ZP}}} \big( \left|\mathcal{F}( z_{\bm{w}}[m])[n] \right| - \Gamma_{\mathrm{thr}} \big)^2,
\end{equation}
where $N_{t}(\alpha)$ is the DFT sample after which no bandwidth is used for a given frequency resolution and $\Gamma_{\mathrm{thr}}$ is a frequency response threshold near zero as explained in Section \ref{subsec:waveforProb}. For example, assuming $\Delta \alpha = 0.1$, we want to generate a waveform for $\alpha=0.3$, it will be $N_{t}(0.3)=14$ as a result of the chosen frequency resolution and the studied roll-off factor. Since the frequency resolution is $\Delta W = 0.1 / (2T)$, the first $11$ samples expand in a bandwidth equal to that for $\alpha = 0$, and $3$ more samples are needed to reach the utilized bandwidth when $\alpha = 0.3$. 
Therefore, for different values of the roll-off factor, $\mathcal{L}_f$ must be adjusted accordingly by $N_{t}(\alpha)$ to capture the used bandwidth of the waveform found by the DFT.  

In addition to equal bandwidth utilization, it is important to ensure that the generated waveform has similar energy to the others for a fair comparison, as illustrated by constraint $\mathrm{C}_1$ in \eqref{eq:optMSE_reform}. The energy of the generated waveform can be calculated using its DFT form and is equal to 
\begin{equation}\label{eq:Pars}
    E_{\mathrm{DNN}} = \frac{1}{N} \sum_{n=0}^{N-1} |\mathcal{F}( z_{\bm{w}}[m])[n]|,
\end{equation}
where $N = N_s +  N_s^{\mathrm{ZP}}$ is the total number of samples in the zero-padded version of the waveform that is extracted through \eqref{eq:DFTcond}. Then, assuming that $E$ is the target energy of the waveform, the following constraint is obtained and passed as part of the loss function:
\begin{equation}\label{eq:energyCond}
    \mathcal{L}_{e} = \frac{1}{|\mathcal{D}|}\sum_{d \in \mathcal{D}} \big( E_{\mathrm{DNN}} - E \big)^2.
\end{equation}

Finally, the generated waveform requires symmetry for its samples w.r.t. the ideal sampling time $t=0$ as indicated by constrained $\mathrm{C}_3$ in \eqref{eq:optMSE_reform}. To achieve this, a constraint is introduced that aims to minimize the distance between the negative and positive parts of the waveform, as follows
\begin{equation}\label{eq:lossSym}
    \mathcal{L}_s = \frac{1}{|\mathcal{D}|}\sum_{d \in \mathcal{D}} \sum_{m=1}^{\frac{N_s - 1}{2}}\big( z_{\bm{w}}[m] - z_{\bm{w}}[-m] \big)^2.
\end{equation}
 
Combining all of the above loss function components described by \eqref{eq:lossMSE}, \eqref{eq:DFTloss}, \eqref{eq:energyCond}, and \eqref{eq:lossSym} yields a total training loss function that satisfies all of the required constraints, aims to minimize the weighted MSE of the received signal for more accurate OTA computing and is given by
\begin{equation}\label{eq:DNNtotloss}
    \mathcal{L}_{tot} = \mathcal{L}_{\mathrm{MSE}} + M_{f}\mathcal{L}_{f} + M_{e}\mathcal{L}_{e} + M_{s}\mathcal{L}_{s}.
\end{equation}
Note that $M_{f}, M_{e}$ and $M_{s}$ act as penalty factors, factors, i.e., violating the corresponding constraint will increase the training loss, so that the training will take the constraint into account and learn to stay within small violation limits. The same reasoning is applied to all considered constraints to ensure that bandwidth, energy, and symmetry violations are kept at extremely low levels, which in turn allows the generation of a waveform that has the same characteristics as RC and BTRC, but outperforms both due to the consideration of MSE in the training phase. It should be highlighted that conventional loss functions cannot be directly used since they do not take into account the required waveform constraints as considered by the custom loss function \eqref{eq:DNNtotloss}.

However, during training, the DNN generates a different waveform for each input sample of the training dataset. Therefore, to generate only one waveform that shows good performance for each input sample of the dataset, during testing only, we generate a waveform by averaging all the generated output waveforms for all input samples. Since the imposed constraints in the loss function have the distributive property, this final waveform also satisfies all constraints. This final waveform is the one we plot in Section \ref{sec:Results} and on which the MSE is calculated. 

Based on the discussion of this section, the proposed DNN waveform framework is suitable to generate band-limited waveforms that satisfy Nyquist's criterion and are symmetric with total energy equal to energy target $E$.

\section{Simulation Results}\label{sec:Results}

In this section, we present the simulation results. For all simulations, we assume that the channel fading follows the circularly symmetric complex Gaussian distribution, i.e., $h_k \sim \mathcal{CN}(0,1),\ \forall k \in \mathcal{K}$, and without loss of generality, the transmitted data are generated from a uniform distribution in the interval $[-\sqrt{3},\sqrt{3}]$. Unless otherwise stated, we assume that $K = 20$ devices participate in the OTA transmission 
We study the problem of MSE minimization for $\mu = 6$ ISI symbols, which means that the time duration of the studied waveforms extends from $-3.5T$ to $3.5T$, and we are interested in generating waveforms that perform better for a roll-off factor step equal to $\Delta \alpha = 0.1$, as an indicator for the effectiveness of the proposed framework while it also provides a satisfactory performance estimation for the whole range of roll-off factors. 

\begin{figure}
    \centering
    \includegraphics[width=0.825\columnwidth]{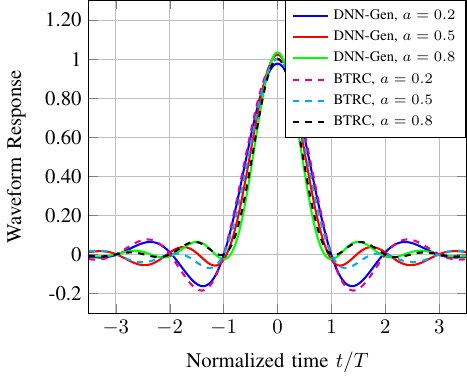}
    \vspace{-3mm}
    \caption{DNN-generated waveforms for OTA computing and various roll-off factors.}
    \label{fig:oddRollOff}
\end{figure}

\subsection{DNN Setup}\label{subsec:dnnParams}
Regarding the DNN setup, the DNN has $3$ hidden layers, each consisting of $256$ neurons, while the ReLU activation function was used. For each simulated roll-off factor of interest, a training dataset of $|\mathcal{D}| = 5 \times 10^4$ channel and data realizations is generated, while the corresponding optimal power allocation vectors were calculated according to \eqref{eq:aMaxISI} and \eqref{eq:powerISI}. The validation and test datasets are generated in a similar manner and consist of $5 \times 10^3$ and $5 \times 10^4$ realizations, respectively. The Adam optimizer was used, while the DNN loss function was defined in \eqref{eq:DNNtotloss}. A batch size of 100 was selected, while the DNN was trained for 10 epochs.

Regarding the parameters associated with \eqref{eq:DNNtotloss}, the number of outputs of the proposed DNN is selected equal to $N_s = 3501$ samples, which allows a time resolution of $\Delta t = T / 500$. To achieve the desired spectral resolution, the zero-padded version of the waveform requires an additional $ N_s^{\mathrm{ZP}} = 6500$ time instances as indicated by \eqref{eq:DFTcond}. For the penalty factors, we selected $M_f = 22.4, M_e = 9000, M_s = 4300$ and for the bandwidth constraint we selected $\Gamma_{\mathrm{thr}} = 0.2$ as a reasonably small frequency response. The selected parameters have been obtained after experimentation to fine-tune the proposed DNN, while the penalty factors are chosen to ensure that the MSE loss remains the dominant part of the loss function, while guaranteeing that any constraint violation is penalized enough by its corresponding factor to ensure minimal constraint violation.

\begin{table}
    \centering
    \caption{Waveform Approximation}
    \label{table:TableCoef}
    \begin{tabular}{|c|c|c|c|}
    \hline
    \multicolumn{4}{|c|}{Approximation Coefficients} \\
    \hline
    \diagbox[width=6em]{Coef.}{$\alpha$} & $\alpha=0.2$ & $\alpha=0.5$ & $\alpha=0.8$ \\
    \hline 
    $a_0$ & $0.0939$ & $0.1313$ & $0.1360$ \\
    \hline
    $a_1$ & $0.2168$ & $0.2638$ & $0.2507$ \\
    \hline
    $a_2$ & $0.1841$ & $0.2371$ & $0.2046$ \\
    \hline
    $a_3$ & $0.2092$ & $0.1676$ & $0.1712$ \\
    \hline
    $a_4$ & $0.1647$ & $0.1406$ & $0.1315$ \\
    \hline
    $a_5$ & $0.0950$ & $0.0764$ & $0.0994$ \\
    \hline
    $a_6$ & $0.0121$ & $0.0053$ & $0.0405$ \\
    \hline
    $p$ & $0.6481$ & $0.8378$ & $0.8739$ \\
    \hline
    RMSE & $3.041 \times 10^{-3}$ & $3.094 \times 10^{-3}$ & $3.132 \times 10^{-3}$ \\
    \hline
    \end{tabular} 
\end{table}

\subsection{Generated Waveforms}\label{subsec:resGenWave}
In Fig. \ref{fig:oddRollOff}, waveforms for various roll-off factors are depicted, each normalized in time. These waveforms exhibit behavior similar to both RC and BTRC waveforms. Although the waveforms resemble one another, the DNN-generated waveforms illustrate better behavior by exhibiting slightly larger decay in the main lobe and having an amplitude close to zero at the ideal sampling time instances $t = qT$ for a larger time interval. This behavior significant impacts system performance, particularly in terms of mitigating sync errors and ISI. For small values of the roll-off factor $\alpha$, the generated waveforms illustrate slower amplitude decay, which does not counter ISI effectively, but has better spectral efficiency due to the smaller utilized bandwidth. Conversely, larger values of $\alpha$ increase the attenuation of ISI, which occurs when symbols in a digital communication system interfere with each other due to channel characteristics. In this scenario, the smaller amplitudes of the sidelobes in the waveforms generated for large $\alpha$ values contribute to the reduction of ISI, thereby improving the system robustness. It is noted that although the maximum values of the generated waveforms are close, they are different. This is a direct consequence of the fact that channel fading is considered during the training phase.

\begin{figure}
    \centering
    \begin{tikzpicture}
        \begin{axis}[
            width=0.825\linewidth,
            xlabel = {DFT Sample},
            ylabel = {Frequency Response},
            ymin = 0,
            ymax = 0.06,
            xmin = 0,
            xmax = 20,
            ytick = {0,0.02,0.04,0.06},
            yticklabels = {0,0.02,0.04,0.06},
            yticklabel style={
                /pgf/number format/fixed,
                /pgf/number format/fixed zerofill,
                /pgf/number format/precision=2,                
            },
            scaled y ticks=false, 
            grid = major,
		legend entries ={{BTRC},{DNN-Gen},{$a=0.2$},{$a=0.5$},{$a=0.8$}},
            legend cell align = {left},
            legend style={font=\footnotesize},
            legend style={at={(1,1)},anchor=north east},
            ]
            \addplot[
            black,
            mark = triangle,
            mark repeat = 1,
            mark size = 2,
            only marks,
            ]
            table {Spectrum_pul/FFT_exp_02.dat};
            \addplot[
            black,
            mark = o,
            mark repeat = 1,
            mark size = 2,
            only marks,
            ]
            table {Spectrum_pul/FFT_final_02.dat};
            \addplot[
            blue,
            mark repeat = 1,
            mark size = 2,
            line width = 1pt,
            style = dotted,
            ]
            table {Spectrum_pul/FFT_final_02.dat};
            \addplot[
            red,
            mark repeat = 1,
            mark size = 2,
            line width = 1pt,
            style = solid,
            ]
            table {Spectrum_pul/FFT_final_05.dat};
            \addplot[
            black,
            mark repeat = 1,
            mark size = 2,
            line width = 1pt,
            style = dashed,
            ]
            table {Spectrum_pul/FFT_final_08.dat};
                        \addplot[
            blue,
            mark = o,
            mark repeat = 1,
            mark size = 2,
            mark options={solid},
            line width = 1pt,
            style = dotted,
            ]
            table {Spectrum_pul/FFT_final_02.dat};
            \addplot[
            red,
            mark = o,
            mark repeat = 1,
            mark size = 2,
            line width = 1pt,
            style = solid,
            ]
            table {Spectrum_pul/FFT_final_05.dat};
            \addplot[
            black,
            mark = o,
            mark repeat = 1,
            mark size = 2,
            mark options={solid},
            line width = 1pt,
            style = dashed,
            ]
            table {Spectrum_pul/FFT_final_08.dat};
            \addplot[
            blue,
            mark = triangle,
            mark repeat = 1,
            mark size = 2,
            mark options={solid},
            line width = 1pt,
            style = dotted,
            ]
            table {Spectrum_pul/FFT_exp_02.dat};
            \addplot[
            red,
            mark = triangle,
            mark repeat = 1,
            mark size = 2,
            mark options={solid},
            line width = 1pt,
            style = solid,
            ]
            table {Spectrum_pul/FFT_exp_05.dat};
            \addplot[
            black,
            mark = triangle,
            mark repeat = 1,
            mark size = 2,
            mark options={solid},
            line width = 1pt,
            style = dashed,
            ]
            table {Spectrum_pul/FFT_exp_08.dat};
        \end{axis}
    \end{tikzpicture}
    \vspace{-3mm}
    \caption{Comparison of bandwidth utilization through DFT.}
    \label{fig:spectrumUtil}
\end{figure}
In Fig. \ref{fig:spectrumUtil}, the bandwidth utilization of the BTRC waveform and the DNN-generated waveforms are presented for some indicative roll-off factors. As can be observed, all waveforms have the same spectrum spread in terms of DFT samples thus, proving the efficiency of the bandwidth-related constraint introduced in the proposed DNN framework through \eqref{eq:DFTloss}. This guarantees that the extracted waveform can be useful in practice by not causing increased intercarrier interference. 

To facilitate the testing of our results without the need to create the DNN architecture, we provide the curve fitting parameters for the DNN-based waveforms of \ref{fig:oddRollOff} in Table \ref{table:TableCoef}. This may also allow practical deployment in real communication systems, since curve fitting can generate waveforms for different roll-off factors, $\alpha$. We use sinusoidal functions for fitting due to the symmetric shape of the curves, thus the generated waveforms can be represented as follows
\begin{equation} \label{eq:generalFit}
    \begin{aligned}
    \hat{z}_{\bm{w}}(t) &= a_0 + a_1 \cos(pt) + a_2 \cos(2pt) + a_3 \cos(3pt) \\
    &\quad + a_4 \cos(4pt) + a_5 \cos(5pt) + a_6 \cos(6pt), 
    \end{aligned}
\end{equation}
with $p$ and $a_{j}, j \in \{ 0, \cdots, 6 \}$ being adjustable coefficients for the approximation.  The root MSE (RMSE) of the fitting approximation is given in Table I as well.

\subsection{Performance Results}\label{subsec:perRes}
In this subsection, we evaluate the performance of the generated waveforms along with the proposed optimal power policies discussed in Section \ref{sec:theoryOpt}, aiming to minimize the MSE.
\subsubsection{OTA under Sync Errors}\label{subsecsec:OTAWithout}

\begin{figure}
    \centering
    \begin{tikzpicture}
        \begin{axis}[
            width=0.825\linewidth,
            xlabel = {Roll-off factor $\alpha$},
            ylabel = {Average MSE},
            ymin = 0.056,
            ymax = 0.064,
            xmin = 0.1,
            xmax = 0.9,
            yticklabel style={
                /pgf/number format/fixed,
                /pgf/number format/fixed zerofill,
                /pgf/number format/precision=3,                
            },
            scaled y ticks=false, 
            grid = major,
		legend entries ={{RC},{BTRC},{DNN-Gen}},
            legend cell align = {left},
            legend style={font=\footnotesize},
            legend style={at={(1,1)},anchor=north east},
            ]
            \addplot[
            blue,
            mark = square,
            mark repeat = 1,
            mark size = 2,
            line width = 1pt,
            style = solid,
            ]
            table {No_ISI/RC_no_ISI_gen.dat};
            \addplot[
            red,
            mark = o,
            mark repeat = 1,
            mark size = 2,
            line width = 1pt,
            style = solid,
            ]
            table {No_ISI/Exp_no_ISI_gen.dat};
            \addplot[
            green,
            mark = triangle,
            mark repeat = 1,
            mark size = 2,
            line width = 1pt,
            style = solid,
            ]
            table {No_ISI/Final_no_ISI_gen.dat};
        \end{axis}
    \end{tikzpicture}
    \vspace{-3mm}
    \caption{Waveform MSE performance for varying roll-off factor and sync errors variance $\sigma_{\epsilon}=0.1$ without the presence of ISI when there are $K=20$ devices.}
    \label{fig:MSE_no_ISI}
\end{figure}
In Fig. \ref{fig:MSE_no_ISI}, we compare the MSE performance of established and DNN-generated waveforms under only the effect of sync errors. Notably, optimal performance for all waveforms is achieved for mid-range roll-off factor values, where the RC waveform performs best. This result is due to the fact that the primary goal of both the BTRC and the DNN-generated waveforms is to mitigate ISI, resulting in inferior performance when only sync errors is present. However, the DNN-generated waveform demonstrates competitive performance across all roll-off factors, even outperforming RC and BTRC for small to medium values of $\alpha$. It is noted that the DNN-based waveform in Fig. 5 was also trained considering ISI.
\subsubsection{OTA under Sync Errors and ISI}\label{subsecsec:OTAWith}

\begin{figure}
    \centering
    \begin{tikzpicture}
        \begin{axis}[
            width=0.825\linewidth,
            xlabel = {Number of devices $K$},
            ylabel = {Average MSE},
            ymin = 0.04,
            ymax = 0.16,
            xmin = 0,
            xmax = 50,
            xtick = {0,10,20,30,40,50},
            xticklabels={0,10,20,30,40,50},
            ytick = {0.04,0.06,0.08,0.1,0.12,0.14,0.16},
            yticklabels = {0.04,0.06,0.08,0.1,0.12,0.14,0.16},
            yticklabel style={
                /pgf/number format/fixed,
                /pgf/number format/fixed zerofill,
                /pgf/number format/precision=2,                
            },
            scaled y ticks=false, 
            grid = both,
            minor grid style={gray!25},
            major grid style={gray!50},
            legend columns=2, 
		legend entries ={{RC},{$a=0.2$},{BTRC},{$a = 0.5$},{DNN-Gen},{$a=0.8$}},
            legend cell align = {left},
            legend style={font=\footnotesize},
            legend style={at={(1,1)},anchor=north east},
            ]
            \addplot[
            black,
            mark = square,
            mark repeat = 1,
            mark size = 2,
            only marks,
            ]
            table {MSE_users/MSE1_rc_02.dat};
            \addplot[
            blue,
            mark = none,
            mark repeat = 1,
            mark size = 2,
            line width = 1pt,
            style = dotted,
            ]
            table {MSE_users/MSE1_rc_02.dat};
            \addplot[
            black,
            mark = triangle,
            mark repeat = 1,
            mark size = 2,
            only marks,
            ]
            table {MSE_users/MSE1_exp_02.dat};
            \addplot[
            red,
            mark = none,
            mark repeat = 1,
            mark size = 2,
            line width = 1pt,
            style = solid,
            ]
            table {MSE_users/MSE1_rc_05.dat};
            \addplot[
            black,
            mark = o,
            mark repeat = 1,
            mark size = 2,
            only marks,
            ]
            table {MSE_users/MSE1_final_02.dat};

            \addplot[
            black,
            mark = none,
            mark repeat = 1,
            mark size = 2,
            line width = 1pt,
            style = dashed,
            ]
            table {MSE_users/MSE1_rc_08.dat};
            \addplot[
            blue,
            mark = square,
            mark repeat = 1,
            mark size = 2,
            mark options={solid},
            line width = 1pt,
            style = dotted,
            ]
            table {MSE_users/MSE1_rc_02.dat};
            \addplot[
            blue,
            mark = triangle,
            mark repeat = 1,
            mark size = 2,
            mark options={solid},
            line width = 1pt,
            style = dotted,
            ]
            table {MSE_users/MSE1_exp_02.dat};
            \addplot[
            blue,
            mark = o,
            mark repeat = 1,
            mark size = 2,
            mark options={solid},
            line width = 1pt,
            style = dotted,
            ]
            table {MSE_users/MSE1_final_02.dat};
            \addplot[
            red,
            mark = square,
            mark repeat = 1,
            mark size = 2,
            line width = 1pt,
            style = solid,
            ]
            table {MSE_users/MSE1_rc_05.dat};
            \addplot[
            black,
            mark = square,
            mark repeat = 1,
            mark size = 2,
            mark options={solid},
            line width = 1pt,
            style = dashed,
            ]
            table {MSE_users/MSE1_rc_08.dat};            
            \addplot[
            red,
            mark = triangle,
            mark repeat = 1,
            mark size = 2,
            line width = 1pt,
            style = solid,
            ]
            table {MSE_users/MSE1_exp_05.dat};
            \addplot[
            black,
            mark = triangle,
            mark repeat = 1,
            mark size = 2,
            mark options={solid},
            line width = 1pt,
            style = dashed,
            ]
            table {MSE_users/MSE1_exp_08.dat};            
            \addplot[
            red,
            mark = o,
            mark repeat = 1,
            mark size = 2,
            line width = 1pt,
            style = solid,
            ]
            table {MSE_users/MSE1_final_05.dat};
            \addplot[
            black,
            mark = o,
            mark repeat = 1,
            mark size = 2,
            mark options={solid},
            line width = 1pt,
            style = dashed,
            ]
            table {MSE_users/MSE1_final_08.dat};
        \end{axis}
    \end{tikzpicture}
    \vspace{-3mm}
    \caption{Waveform MSE performance for varying roll-off factor and sync errors with $\sigma_{\epsilon}=0.1$.}
    \label{fig:MSE_sigma_01}
\end{figure}

Figs. \ref{fig:MSE_sigma_01} and \ref{fig:MSE_sigma_02} illustrate the average MSE for varying numbers of devices in the system. It's evident that employing the DNN-generated waveform offers satisfactory performance compared to the established RC and BTRC waveforms. Notably, this advantage remains consistent across all selected roll-off factors, particularly for mid-range values of $\alpha$, which are commonly used in practice. Moreover, despite the decreased effect of ISI when increasing the roll-off factor, the choice of waveform remains significant. Remarkably, the DNN-generated waveform outperforms the RC waveform, despite the latter utilizing a $30\%$ larger bandwidth. It should also be highlighted that the observed gains in MSE are on the order of $10^{-2}$, whereas conventional waveforms typically exhibit smaller gains when used for traditional communication systems, with BER gains between $10^{-4}$ and $10^{-6}$ for the same amount of synchronization errors.
Moreover, for larger sync errors value of $\sigma_{\epsilon}=0.2$, the improvement provided by the DNN-generated waveform becomes pronounced, aiding in the convergence of the MSE curve. In addition, for small values of $\alpha$, the strong effect of ISI leads the MSE to increase regardless of the waveform. Thus, achieving optimal performance requires considering both roll-off factor and waveform selection to reduce the MSE.

\begin{figure}
    \centering
    \begin{tikzpicture}
        \begin{axis}[
            width=0.825\linewidth,
            xlabel = {Number of devices $K$},
            ylabel = {Average MSE},
            ymin = 0.05,
            ymax = 0.31,
            xmin = 0,
            xmax = 50,
            xtick = {0,10,20,30,40,50},
            xticklabels={0,10,20,30,40,50},
            ytick = {0.05,0.1,...,0.3},
            yticklabels = {0.05,0.1,0.15,0.2,0.25,0.3},
            grid = both,
            minor grid style={gray!25},
            major grid style={gray!50},
            legend columns=3, 
		legend entries ={{RC},{BTRC},{DNN-Gen},{$a=0.2$},{$a=0.5$},{$a=0.8$}},
            legend cell align = {left},
            legend style={font=\footnotesize},
            legend style={at={(1,1)},anchor=north east},
            ]
            \addplot[
            black,
            mark = square,
            mark repeat = 1,
            mark size = 2,
            only marks,
            ]
            table {MSE_users_2/MSE2_rc_02.dat};
            \addplot[
            black,
            mark = triangle,
            mark repeat = 1,
            mark size = 2,
            only marks,
            ]
            table {MSE_users_2/MSE2_exp_02.dat};
            \addplot[
            black,
            mark = o,
            mark repeat = 1,
            mark size = 2,
            only marks,
            ]
            table {MSE_users_2/MSE2_final_02.dat};
            \addplot[
            blue,
            mark = none,
            mark repeat = 1,
            mark size = 2,
            line width = 1pt,
            style = dotted,
            ]
            table {MSE_users_2/MSE2_rc_02.dat};
            \addplot[
            red,
            mark = none,
            mark repeat = 1,
            mark size = 2,
            line width = 1pt,
            style = solid,
            ]
            table {MSE_users_2/MSE2_rc_05.dat};
            \addplot[
            black,
            mark = none,
            mark repeat = 1,
            mark size = 2,
            line width = 1pt,
            style = dashed,
            ]
            table {MSE_users_2/MSE2_rc_08.dat};
            \addplot[
            blue,
            mark = square,
            mark repeat = 1,
            mark size = 2,
            mark options={solid},
            line width = 1pt,
            style = dotted,
            ]
            table {MSE_users_2/MSE2_rc_02.dat};
            \addplot[
            blue,
            mark = triangle,
            mark repeat = 1,
            mark size = 2,
            mark options={solid},
            line width = 1pt,
            style = dotted,
            ]
            table {MSE_users_2/MSE2_exp_02.dat};
            \addplot[
            blue,
            mark = o,
            mark repeat = 1,
            mark size = 2,
            mark options={solid},
            line width = 1pt,
            style = dotted,
            ]
            table {MSE_users_2/MSE2_final_02.dat};
            \addplot[
            red,
            mark = square,
            mark repeat = 1,
            mark size = 2,
            line width = 1pt,
            style = solid,
            ]
            table {MSE_users_2/MSE2_rc_05.dat};
            \addplot[
            black,
            mark = square,
            mark repeat = 1,
            mark size = 2,
            mark options={solid},
            line width = 1pt,
            style = dashed,
            ]
            table {MSE_users_2/MSE2_rc_08.dat};            
            \addplot[
            red,
            mark = triangle,
            mark repeat = 1,
            mark size = 2,
            line width = 1pt,
            style = solid,
            ]
            table {MSE_users_2/MSE2_exp_05.dat};
            \addplot[
            black,
            mark = triangle,
            mark repeat = 1,
            mark size = 2,
            mark options={solid},
            line width = 1pt,
            style = dashed,
            ]
            table {MSE_users_2/MSE2_exp_08.dat};            
            \addplot[
            red,
            mark = o,
            mark repeat = 1,
            mark size = 2,
            line width = 1pt,
            style = solid,
            ]
            table {MSE_users_2/MSE2_final_05.dat};
            \addplot[
            black,
            mark = o,
            mark repeat = 1,
            mark size = 2,
            mark options={solid},
            line width = 1pt,
            style = dashed,
            ]
            table {MSE_users_2/MSE2_final_08.dat};
        \end{axis}
    \end{tikzpicture}
    \vspace{-3mm}
    \caption{Waveform MSE performance for varying roll-off factor and sync errors with $\sigma_{\epsilon}=0.2$.}
    \label{fig:MSE_sigma_02}
\end{figure}

\begin{table}
    \centering
    \caption{MSE Improvement Table}
    \label{table:TableMSE}
    \begin{tabular}{|c|c|c|c|c|}
    \hline
    \multicolumn{5}{|c|}{MSE Performance Gain ($\%$)} \\
    \hline
    & \multicolumn{2}{c|}{$\sigma_{\epsilon} = 0.1$} & \multicolumn{2}{c|}{$\sigma_{\epsilon} = 0.2$} \\
    \hline
    \diagbox[width=12em]{\textbf{Roll-off Factor}}{\textbf{Waveform}} & \textbf{BTRC} & \textbf{RC} & \textbf{BTRC} & \textbf{RC}\\
    \hline 
    $\alpha=0.1$ & $2.16$ & $4.27$ & $4.09$ & $6$ \\
    \hline
    $\alpha=0.2$ & $5.65$ & $12.39$ & $7.20$ & $13.38$\\
    \hline
    $\alpha=0.3$ & $3.38$ & $13.14$ & $9.45$ & $19.95$\\
    \hline
    $\alpha=0.4$ & $5.95$ & $15.96$ & $12$ & $25.31$\\
    \hline
    $\alpha=0.5$ & $8.82$ & $21.06$ & $14.54$ & $29.56$\\
    \hline
    $\alpha=0.6$ & $5.62$ & $19.68$ & $13.15$ & $30.97$\\
    \hline
    $\alpha=0.7$ & $5.78$ & $18.66$ & $12.88$ & $31.31$\\
    \hline
    $\alpha=0.8$ & $2.49$ & $15.73$ & $9.36$ & $29.08$\\
    \hline
    $\alpha=0.9$ & $1.76$ & $12.58$ & $7.59$ & $26.14$\\
    \hline
    $\alpha=1.0$ & $2.16$ & $8.29$ & $4.69$ & $20.84$\\
    \hline  
    \end{tabular} 
\end{table}

Table \ref{table:TableMSE} presents the percentage improvement gain of DNN-generated waveforms over their RC and BTRC counterparts for various roll-off factors and sync errors. The DNN-based waveforms demonstrate significant gains, particularly for mid-range roll-off factors. This is reasonable as waveforms in this range retain relatively large sidelobe amplitudes, resulting in a greater impact of ISI on the average MSE. Conversely, for large roll-off factors, the ISI effect diminishes due to the waveform having smaller sidelobes, while the DNN's ability to generate higher-gain waveforms is constrained by bandwidth resolution limitations for small roll-off factors. Nonetheless, in practical applications where spectrum allocation matters, mid-range roll-off factors are common. In such cases, DNN-generated waveforms exhibit significant performance gains, making them a good alternative to established waveforms.


\begin{figure}
    \centering
    \begin{tikzpicture}
        \begin{axis}[
            width=0.825\linewidth,
            xlabel = {Transmit SNR},
            ylabel = {Average MSE},
            ymin = 0,
            ymax = 0.25,
            xmin = 0,
            xmax = 20,
            ytick = {0,0.05,0.1,0.15,0.2,0.25},
            yticklabels = {0,0.05,0.1,0.15,0.2,0.25},
            scaled y ticks=false, 
            grid = both,
            minor grid style={gray!25},
            major grid style={gray!50},
            legend columns=3, 
		legend entries ={{RC},{BTRC},{DNN-Gen},{$a=0.2$},{$a=0.5$},{$a=0.8$}},
            legend cell align = {left},
            legend style={font=\footnotesize},
            legend style={at={(1,1)},anchor=north east},
            ]
            \addplot[
            black,
            mark = square,
            mark repeat = 1,
            mark size = 2,
            only marks,
            ]
            table {MSE_P/MSE1_P_rc_02.dat};
            \addplot[
            black,
            mark = triangle,
            mark repeat = 1,
            mark size = 2,
            only marks,
            ]
            table {MSE_P/MSE1_P_exp_02.dat};
            \addplot[
            black,
            mark = o,
            mark repeat = 1,
            mark size = 2,
            only marks,
            ]
            table {MSE_P/MSE1_P_final_02.dat};
            \addplot[
            blue,
            mark = none,
            mark repeat = 1,
            mark size = 2,
            line width = 1pt,
            style = dotted,
            ]
            table {MSE_P/MSE1_P_rc_02.dat};
            \addplot[
            red,
            mark = none,
            mark repeat = 1,
            mark size = 2,
            line width = 1pt,
            style = solid,
            ]
            table {MSE_P/MSE1_P_rc_05.dat};
            \addplot[
            black,
            mark = none,
            mark repeat = 1,
            mark size = 2,
            line width = 1pt,
            style = dashed,
            ]
            table {MSE_P/MSE1_P_rc_08.dat};
            \addplot[
            blue,
            mark = square,
            mark repeat = 1,
            mark size = 2,
            mark options={solid},
            line width = 1pt,
            style = dotted,
            ]
            table {MSE_P/MSE1_P_rc_02.dat};
            \addplot[
            blue,
            mark = triangle,
            mark repeat = 1,
            mark size = 2,
            mark options={solid},
            line width = 1pt,
            style = dotted,
            ]
            table {MSE_P/MSE1_P_exp_02.dat};
            \addplot[
            blue,
            mark = o,
            mark repeat = 1,
            mark size = 2,
            mark options={solid},
            line width = 1pt,
            style = dotted,
            ]
            table {MSE_P/MSE1_P_final_02.dat};
            \addplot[
            red,
            mark = square,
            mark repeat = 1,
            mark size = 2,
            line width = 1pt,
            style = solid,
            ]
            table {MSE_P/MSE1_P_rc_05.dat};
            \addplot[
            black,
            mark = square,
            mark repeat = 1,
            mark size = 2,
            mark options={solid},
            line width = 1pt,
            style = dashed,
            ]
            table {MSE_P/MSE1_P_rc_08.dat};            
            \addplot[
            red,
            mark = triangle,
            mark repeat = 1,
            mark size = 2,
            line width = 1pt,
            style = solid,
            ]
            table {MSE_P/MSE1_P_exp_05.dat};
            \addplot[
            black,
            mark = triangle,
            mark repeat = 1,
            mark size = 2,
            mark options={solid},
            line width = 1pt,
            style = dashed,
            ]
            table {MSE_P/MSE1_P_exp_08.dat};            
            \addplot[
            red,
            mark = o,
            mark repeat = 1,
            mark size = 2,
            line width = 1pt,
            style = solid,
            ]
            table {MSE_P/MSE1_P_final_05.dat};
            \addplot[
            black,
            mark = o,
            mark repeat = 1,
            mark size = 2,
            mark options={solid},
            line width = 1pt,
            style = dashed,
            ]
            table {MSE_P/MSE1_P_final_08.dat};
        \end{axis}
    \end{tikzpicture}
    \vspace{-3mm}
    \caption{Waveform MSE performance for varying roll-off factor and sync errors with $\sigma_{\epsilon}=0.1$.}
    \vspace{-3mm}
    \label{fig:MSE_P_sigma_01}
\end{figure}

In Fig. \ref{fig:MSE_P_sigma_01}, we examine the MSE performance as the transmit SNR varies. Typically, lower SNRs result in larger MSE due to significant noise effect in addition to ISI. However, we observe that as the SNR increases, the MSE decreases, with the DNN-generated waveform consistently outperforming both RC and BTRC across the entire SNR range of 0 to 20 dB. This demonstrates the robustness of the proposed approach to changes in transmit power and noise levels. In addition, as shown, the DNN-generated waveform appears to achieve $2-3$ dB gain over BTRC and more than $5$ dB over RC for intermediate SNR values, which is a significant gain, while these gains are even more pronounced for larger sync error variance, as highlighted in Table \ref{table:TableMSE}.

\begin{figure}
    \centering
    \begin{tikzpicture}
        \begin{axis}[
            width=0.825\linewidth,
            xlabel = {Roll-off factor $\alpha$},
            ylabel = {Average MSE},
            ymin = 0.05,
            ymax = 0.12,
            xmin = 0.2,
            xmax = 0.8,
            yticklabel style={
                /pgf/number format/fixed,
                /pgf/number format/fixed zerofill,
                /pgf/number format/precision=2,                
            },
            scaled y ticks=false, 
            grid = major,
		legend entries ={{Gaussian},{Uniform},{Laplacian},{RC},{BTRC},{DNN-Gen}},
            legend cell align = {left},
            legend style={font=\footnotesize},
            legend style={at={(1,1)},anchor=north east},
            ]
            \addplot[
            black,
            mark = square,
            mark repeat = 1,
            mark size = 2.5,
            line width = 1pt,
            style = solid,
            ]
            table {Gaussian/Gaussian_MSE_full.dat};
            \addplot[
            black,
            mark = triangle,
            mark repeat = 1,
            mark size = 2.5,
            line width = 1pt,
            style = solid,
            ]
            table {Uniform/Uniform_MSE_full.dat};
            \addplot[
            black,
            mark = o,
            mark repeat = 1,
            mark size = 2.5,
            line width = 1pt,
            style = solid,
            ]
            table {Laplace/MSE_lap_final.dat};
            \addplot[
            blue,
            mark = none,
            mark repeat = 1,
            mark size = 2.5,
            line width = 1pt,
            style = dotted,
            ]
            table {Gaussian/Gaussian_MSE_RC.dat};
            \addplot[
            red,
            mark = none,
            mark repeat = 1,
            mark size = 2.5,
            mark options={solid},
            line width = 1pt,
            style = dashed,
            ]
            table {Gaussian/Gaussian_MSE_Exp.dat};
            \addplot[
            green,
            mark = none,
            mark repeat = 1,
            mark size = 2.5,
            line width = 1pt,
            style = solid,
            ]
            table {Gaussian/Gaussian_MSE_full.dat}; 
            \addplot[
            green,
            mark = square,
            mark repeat = 1,
            mark size = 2.5,
            line width = 1pt,
            style = solid,
            ]
            table {Gaussian/Gaussian_MSE_full.dat};
            \addplot[
            red,
            mark = square,
            mark repeat = 1,
            mark size = 2.5,
            mark options={solid},
            line width = 1pt,
            style = dashed,
            ]
            table {Gaussian/Gaussian_MSE_Exp.dat};
            \addplot[
            blue,
            mark = square,
            mark repeat = 1,
            mark size = 2.5,
            mark options={solid},
            line width = 1pt,
            style = dotted,
            ]
            table {Gaussian/Gaussian_MSE_RC.dat};
            \addplot[
            green,
            mark = triangle,
            mark repeat = 1,
            mark size = 2.5,
            line width = 1pt,
            style = solid,
            ]
            table {Uniform/Uniform_MSE_full.dat};
            \addplot[
            red,
            mark = triangle,
            mark repeat = 1,
            mark size = 2.5,
            mark options={solid},
            line width = 1pt,
            style = dashed,
            ]
            table {Uniform/Uniform_MSE_Exp.dat};
            \addplot[
            blue,
            mark = triangle,
            mark repeat = 1,
            mark size = 2.5,
            mark options={solid},
            line width = 1pt,
            style = dotted,
            ]
            table {Uniform/Uniform_MSE_RC.dat};
            \addplot[
            green,
            mark = o,
            mark repeat = 1,
            mark size = 2.5,
            line width = 1pt,
            style = solid,
            ]
            table {Laplace/MSE_lap_final.dat};
            \addplot[
            red,
            mark = o,
            mark repeat = 1,
            mark size = 2.5,
            mark options={solid},
            line width = 1pt,
            style = dashed,
            ]
            table {Laplace/MSE_lap_Exp.dat};
            \addplot[
            blue,
            mark = o,
            mark repeat = 1,
            mark size = 2.5,
            mark options={solid},
            line width = 1pt,
            style = dotted,
            ]
            table {Laplace/MSE_lap_RC.dat};
        \end{axis}
    \end{tikzpicture}
    \vspace{-3mm}
    \caption{Waveform MSE performance for various data distributions (Gaussian, Laplacian, Uniform) and sync errors with $\sigma_{\epsilon} = 0.1 $.}
    \label{fig:GaussData}
\end{figure}

Fig. \ref{fig:GaussData} shows the effect of different data distributions on the average MSE with Gaussian distribution, $x_k \sim \mathcal{N}(0,1), \forall k \in \mathcal{K}$, and Laplacian distribution, $x_k \sim \mathrm{Laplace}(0,1/\sqrt{2}), \forall k \in \mathcal{K}$, simulated. The distributions' parameters selection are such that zero mean and unit variance holds for fair comparison among all distributions. As can be seen, the performance remains identical for all plotted values $\alpha$ for the conventional waveforms (RC and BTRC) and the DNN-generated one. This behavior highlights the independence of the proposed framework from the input data, as explained in Section \ref{sec:DNNWaveforms}, thus proving its generality for different device data distributions. Furthermore, the DNN-generated waveform outperforms RC and BTRC for all roll-off factors.

\begin{figure}[ht!]
    \centering
    \begin{tikzpicture}
        \begin{axis}[
            width=0.825\linewidth,
            xlabel = {Roll-off factor $\alpha$},
            ylabel = {Average MSE},
            ymin = 0.05,
            ymax = 0.2,
            xmin = 0.2,
            xmax = 0.8,
            yticklabel style={
                /pgf/number format/fixed,
                /pgf/number format/fixed zerofill,
                /pgf/number format/precision=2,                
            },
            scaled y ticks=false, 
            grid = major,
		legend entries ={{Perfect CSI},{Imperfect CSI},{RC},{BTRC},{DNN-Gen}},
            legend cell align = {left},
            legend style={font=\footnotesize},
            legend style={at={(1,1)},anchor=north east},
            ]
            \addplot[
            black,
            mark = square,
            mark repeat = 1,
            mark size = 2.5,
            line width = 1pt,
            style = solid,
            ]
            table {Uniform/Uniform_MSE_full.dat};
            \addplot[
            black,
            mark = triangle,
            mark repeat = 1,
            mark size = 2.5,
            mark options={solid},
            line width = 1pt,
            style = solid,
            ]                     table{MSE_imperfect_CSI/MSE_imperfect_final.dat};
            \addplot[
            blue,
            mark = none,
            mark repeat = 1,
            mark size = 2.5,
            line width = 1pt,
            style = dotted,
            ]
            table {Uniform/Uniform_MSE_RC.dat};
            \addplot[
            red,
            mark = none,
            mark repeat = 1,
            mark size = 2.5,
            line width = 1pt,
            style = dashed,
            ]
            table {Uniform/Uniform_MSE_Exp.dat};
            \addplot[
            green,
            mark = none,
            mark repeat = 1,
            mark size = 2.5,
            line width = 1pt,
            style = solid,
            ]
            table {Uniform/Uniform_MSE_full.dat}; 
            \addplot[
            green,
            mark = square,
            mark repeat = 1,
            mark size = 2.5,
            line width = 1pt,
            style = solid,
            ]
            table {Uniform/Uniform_MSE_full.dat};
            \addplot[
            red,
            mark = square,
            mark repeat = 1,
            mark size = 2.5,
            mark options={solid},
            line width = 1pt,
            style = dashed,
            ]
            table {Uniform/Uniform_MSE_Exp.dat};
            \addplot[
            blue,
            mark = square,
            mark repeat = 1,
            mark size = 2.5,
            mark options={solid},
            line width = 1pt,
            style = dotted,
            ]
            table {Uniform/Uniform_MSE_RC.dat};
            \addplot[
            green,
            mark = triangle,
            mark repeat = 1,
            mark size = 2.5,
            mark options={solid},
            line width = 1pt,
            style = solid,
            ]
            table {MSE_imperfect_CSI/MSE_imperfect_final.dat};
            \addplot[
            red,
            mark = triangle,
            mark repeat = 1,
            mark size = 2.5,
            mark options={solid},
            line width = 1pt,
            style = dashed,
            ]
            table {MSE_imperfect_CSI/MSE_imperfect_Exp.dat};
            \addplot[
            blue,
            mark = triangle,
            mark repeat = 1,
            mark size = 2.5,
            mark options={solid},
            line width = 1pt,
            style = dotted,
            ]
            table {MSE_imperfect_CSI/MSE_imperfect_RC.dat};
        \end{axis}
    \end{tikzpicture}
    
    \vspace{-3mm}
    \caption{MSE performance for perfect and imperfect CSI conditions and sync errors with $\sigma_{\epsilon}=0.1$.}
    \label{fig:imperfectCSI}
\end{figure}

To investigate the impact of imperfect CSI knowledge simulations for the performance of all waveforms are present in \ref{fig:imperfectCSI}. As such, an error $e_k \sim \mathcal{CN}(0,\sigma^2/P)$ that is assumed to be caused by noise during the channel estimation exists, so that the channel estimation is given as $h_k' = h_k + e_k$. As observed, the performance of the DNN-generated waveform outperforms both RC and BTRC waveforms in this practical scenario as well, exhibiting even larger gains. This characteristic is attributed to the fact that imperfect CSI mainly introduces additional time shifts, while the magnitude does not cause such a great degradation. These additional time shifts can be considered as a larger sync error at each device, and since the generated waveform is able to mitigate even larger sync errors, as observed by Fig. \ref{fig:MSE_sigma_02} and Table \ref{table:TableMSE}, it can counter imperfect CSI considerably better than the conventional waveforms proving the increased robustness of the DNN-generated waveform to imperfect CSI.


Finally, another important observation drawn from the presented diagrams relates to the magnitude of the average MSE caused by ISI. Comparing these figures, specifically when there are $K=20$ devices and the transmit SNR is 10 dB, it is visible that the average MSE is considerably greater (around $60-70\%$) when ISI occurs. Also, under ISI the average MSE is heavily dependent on the choice of the roll-off factor, while under only sync errors, the MSE is almost the same for all values of $\alpha$.

\section{Conclusions}\label{sec:conclusions}
Our study provides a practical examination of OTA computing performance considering sync errors and ISI. We derived the theoretical MSE for the OTA transmission and established optimal power allocation strategies to minimize the MSE. Additionally, we introduced a novel DNN-based method for waveform design, integrating power, bandwidth, and design constraints into the DNN loss function. Simulation results confirmed our theoretical findings and demonstrated the superior performance of the designed waveform compared to the traditional RC and BTRC waveforms. As future extensions of this work, the introduction of MIMO for waveform design could be interesting as well as addressing imperfect CSI or improving the current DNN-based approach. An area for improvement lies in our method of obtaining a single waveform from the DNN output, which currently involves averaging multiple output waveforms. 
\appendices
\section{Proof of Proposition \ref{lemma:lemma1}} \label{app:lemma1}
It is known that for the moments of a Gaussian random variable $X \sim \mathcal{N}(0, \sigma^2)$ it holds that 
\begin{equation}\label{eq:MomProperty}
    \mathbb{E}_{X}[X^{m}] = 
    \begin{cases}
        0, & \text{if } m \text{ is odd} \\ 
        \sigma^{m}(m-1)!!, & \text{if } m \text{ is even} ,
    \end{cases}
\end{equation}
where $(m-1)!! = 1 \cdots (m-3)(m-1)$ is the double factorial.

Let $g(\epsilon) = \mathrm{sinc}(\epsilon)\cos(\pi\alpha \epsilon)$ be the raised cosine waveform without the denominator term, $1- (2a\epsilon)^2$, in \eqref{eq:RC} which is strongly approximated by \eqref{eq:approxFrac}. Using the Taylor expansion of the $\mathrm{sinc}$ and $\cos$ functions, we can write
\begin{equation}\label{eq:RCTaylor}
    \begin{aligned}
    g(\epsilon) &= \sum_{n=0}^{\infty} \frac{(-1)^{n}(\pi \epsilon)^{2n}}{(2n+1)!} \sum_{k=0}^{\infty} \frac{(-1)^{k}(\pi \alpha \epsilon)^{2k}}{2k!} \\
    &= \sum_{m=0}^{\infty} \underbrace{\sum_{n=0}^{m} \frac{(-1)^{m}\pi^{2m}\alpha^{2(m-n)}}{(2n+1)!(2m-2n)!}}_{\kappa_m} \epsilon^{2m},
    \end{aligned}
\end{equation} 
where $m=n+k$. With the same technique, we can also write
\begin{align}\label{eq:RCSqTaylor}
    g^2(\epsilon) &= \sum_{n=0}^{\infty} \frac{(-1)^{n}(\pi \epsilon)^{2n}}{(2n+1)!} \sum_{m=0}^{\infty} \frac{(-1)^{m}(\pi \epsilon)^{2m}}{(2m+1)!}  \nonumber \\
    &\quad \times  \sum_{k=0}^{\infty} \frac{(-1)^{k}(\pi \alpha \epsilon)^{2k}}{2k!} \sum_{l=0}^{\infty} \frac{(-1)^{l}(\pi \alpha \epsilon)^{2l}}{2l!}  \\
    &= \sum_{p=0}^{\infty} \underbrace{ \sum_{n=0}^{p} \sum_{l=0}^{p-n} \sum_{k=0}^{p-n-l} \hspace{-2mm}\tfrac{(-1)^{p}\pi^{2p}\alpha^{2(p-l-n)}}{(2n+1)!(2l+1)!(2k)!(2(p-n-k-l))!}}_{\lambda_p} \epsilon^{2p} \nonumber
\end{align}
with $p = n+m+k+l$. Then, using \eqref{eq:MomProperty}, \eqref{eq:RCTaylor} and \eqref{eq:approxFrac}, for the mean value of the waveform amplitude $\bar{\epsilon}_{1} = \mathbb{E}_{\epsilon}[g(\epsilon)(1+4\alpha^2\epsilon^2)]$, it holds that
\begin{equation}\label{eq:sincEproof}
    \bar{\epsilon}_{1} = \sum_{m=0}^{\infty} \! \kappa_{m} \! \left( \! \frac{1}{\sqrt{2\pi \sigma_{\epsilon}^2}}\int_{-\infty}^{\infty} \epsilon^{2m}(1 + 4 \alpha^2 \epsilon^2) e^{-\frac{\epsilon^2}{2\sigma_{\epsilon}^2}} \mathrm{d} \epsilon \right)
\end{equation}
and \eqref{eq:sincE} follows. Similarly, using \eqref{eq:MomProperty}, \eqref{eq:RCSqTaylor} and \eqref{eq:approxFrac}, for the mean squared amplitude $\bar{\epsilon}_{2} = \mathbb{E}_{\epsilon}[g^2(\epsilon)(1+4\alpha^2\epsilon^2)^2]$, we obtain
\begin{equation}\label{eq:sincSqEproof}
    \bar{\epsilon}_{2} = \sum_{p=0}^{\infty} \lambda_{p} \left( \frac{1}{\sqrt{2\pi \sigma_{\epsilon}^2}}\int_{-\infty}^{\infty} \epsilon^{2p}(1 + 4 \alpha^2 \epsilon^2)^2 e^{-\frac{\epsilon^2}{2\sigma_{\epsilon}^2}} \mathrm{d} \epsilon \right)
\end{equation}
and \eqref{eq:sincESq} follows, which holds for any device $k \in \mathcal{K}$, and the proof is completed.

\section{Proof of Proposition \ref{lemma:lemma2}}\label{app:lemma2}
To compute the mean squared amplitude at $t=T(1+\epsilon)$, the procedure is similar to that in Proposition \ref{lemma:lemma1} by expanding the $\mathrm{sinc}$ function and using the binomial expansion as 
\begin{equation}\label{eq:sinc}
    \begin{aligned}
    \mathrm{sinc}(1+\epsilon) &= \sum_{n=0}^{\infty} \frac{(-1)^{n}\pi^{2n}}{(2n+1)!} {(1+\epsilon)}^{2n}   \\
    &= \sum_{n=0}^{\infty} \frac{(-1)^{n}\pi^{2n}}{(2n+1)!}  \sum_{k=0}^{2n} \binom{2n}{k} \epsilon^{k}  \\
    &= 
    \sum\limits_{\substack{k=1, \\
        k \text{ odd}}}^{\infty} \underbrace{ \sum\limits_{n=\frac{k+1}{2}}^{\infty} \hspace{-2mm}\frac{(-1)^n\pi ^{2n}}{(2n+1)!}\binom{2n}{k}}_{A_k}  \epsilon^{k}
         \\
     & \quad +   \sum\limits_{\substack{k=0, \\
        k \text{ even} }}^{\infty}\underbrace{\sum\limits_{n=\frac{k}{2}}^{\infty} \frac{(-1)^n\pi ^{2n}}{(2n+1)!}\binom{2n}{k}}_{B_k}\epsilon^{k},
    \end{aligned}
\end{equation}
where $A_k = 0$ whenever $k$ is even and $B_k = 0$ whenever $k$ is odd, respectively. 
Then, according to \eqref{eq:MomProperty}, for the mean value of $\mathrm{sinc}^2(1+\epsilon)$ to be non-zero there are two possible combinations, i.e., odd with odd terms and even with even terms. Therefore, it holds that
\begin{equation}\label{eq:sincSq}
    \begin{aligned}
    \mathbb{E}_{\epsilon}[\mathrm{sinc}^2(1+\epsilon)] &= 
        \mathbb{E}_{\epsilon} \left[  \sum\limits_{\substack{k=1, \\
        k \text{ odd}}}^{\infty} A_k \epsilon^{k} \sum\limits_{\substack{l=1, \\
        l \text{ odd}}}^{\infty} A_l \epsilon^{l}\right] \\
        & \quad + \mathbb{E}_{\epsilon} \left[
        \sum\limits_{\substack{k=0, \\
        k \text{ even}}}^{\infty} B_k \epsilon^{k} \sum\limits_{\substack{l=0, \\
        l \text{ even}}}^{\infty} B_l \epsilon^{l} \right]
    \\
    &= 
    \mathbb{E}_{\epsilon} \left[ \sum\limits_{\substack{m=2, \\
        m \text{ even}}}^{\infty} \sum\limits_{\substack{n=1, \\
        n \text{ odd}}}^{m} A_n A_{m-n} \epsilon^{m} \right]
         \\
     & \quad +   \mathbb{E}_{\epsilon} \left[ \sum\limits_{\substack{m=0, \\
        m \text{ even}}}^{\infty}  \sum\limits_{\substack{n=0, \\
        n \text{ even}}}^{m} B_n B_{m-n}  \epsilon^{m} \right],
    \end{aligned}
\end{equation}
which can be equivalently reduced to 
\begin{equation}
    \tilde{\epsilon}_{1} \hspace{-0.8mm} = \hspace{-0.8mm} \mathbb{E}_{\epsilon}[\mathrm{sinc}^2(1\hspace{-0.5mm}+\hspace{-0.5mm}\epsilon)] \hspace{-1mm} = \hspace{-1mm} \mathbb{E}_{\epsilon} \hspace{-1mm} \left[ \sum\limits_{m=0}^{\infty} \sum\limits_{n=0}^{2m} \hspace{-0.2mm} A_{n}A_{2m-n} \hspace{-1mm} +  \hspace{-1mm} B_n B_{2m-n} \epsilon^{2m} \hspace{-0.5mm} \right] \hspace{-1mm},
\end{equation}
from which we can conclude \eqref{eq:alpha0} using \eqref{eq:MomProperty} and $B_0^2 = 1$.

Then, due to the symmetry of the Gaussian random variable with zero mean, it holds that
\begin{equation}\label{eq:sincGausSym}
    \mathbb{E}_{\epsilon}[\mathrm{sinc}^2(-\hspace{-0.3mm}1\hspace{-0.3mm}+\hspace{-0.2mm}\epsilon)] = \mathbb{E}_{\epsilon}[\mathrm{sinc}^2(-\hspace{-0.3mm}1\hspace{-0.3mm}-\hspace{-0.2mm}\epsilon)] = \mathbb{E}_{\epsilon}[\mathrm{sinc}^2(1\hspace{-0.3mm}+\hspace{-0.25mm}\epsilon)],
\end{equation}
where the second equality holds due to the fact that $\mathrm{sinc}^2(\cdot)$ is an even function. Thus, $\tilde{\epsilon}_{1} = \tilde{\epsilon}_{-1}$, which holds for any device $k \in \mathcal{K}$ and the proof is completed.

\section{Proof of Proposition \ref{prop:equiv}}\label{app:lemma3}
Assuming that one receiver sync error $\epsilon$ exists, the received signal can be expressed as
\begin{equation}\label{eq:RecISI_r}
        \begin{aligned}
    [\hat{y}_{\mathrm{ISI}}]^{(r)} &= a \left( \sum_{k=1}^{K}x_{k,0}z_k(\epsilon T) b_kh_{k} + n \right) \\
    & \quad + a\underbrace{ \left( \sum_{\substack{q = - \frac{\mu}{2}  \\
    q \neq 0}}^{ \frac{\mu}{2} } \sum_{k=1}^{K}x_{k,q}z_k(T(q + \epsilon))b_kh_{k} \right)}_{\text{ISI terms}}.
    \end{aligned}
\end{equation}
Leveraging the independence of the transmitted data at different time instances, resulting in $\mathbb{E}_{x_k}[x_{k,q_{1}}x_{k,q_{2}}] = 0$, when $q_{1} \neq q_{2}$, the MSE w.r.t. the noise $n$ and $x_k, \forall k \in \mathcal{K}$, is obtained as
\begin{equation}\label{eq:recFullISI_r}
    \begin{aligned}
    \mathrm{\overline{MSE}}^{\mathrm{ISI}}(a,\bm{b}) &= \sum_{k=1}^{K} \left( az_k(\epsilon T)b_kh_k - 1 \right)^2 + \sigma^2a^2 \\
    & \quad + \sum_{\substack{q = -  \frac{\mu}{2}  \\
    q \neq 0}}^{ \frac{\mu}{2} } \sum_{k=1}^{K} \left( az_k(T(q+\epsilon))b_kh_{k} \right)^2.
    \end{aligned}
\end{equation}

Taking the expectation of \eqref{eq:recFullISI_r} w.r.t. $\epsilon$ to get $\mathrm{MSE}^{\mathrm{ISI}}(a,\bm{b}) = \mathbb{E}_{\epsilon} \left[ \mathrm{\overline{MSE}}^{\mathrm{ISI}}(a,\bm{b})\right]$ and making use of the linearity of the mean value yields only terms of the form $\tilde{\epsilon}_{q} =\mathbb{E}_{\epsilon}[z_k^2(T(q+\epsilon))]$ and $\check{\epsilon} = \mathbb{E}_{\epsilon}[z_k(\epsilon T)]$ as
\begin{align}
    \left[\mathrm{MSE}^{\mathrm{ISI}}(a,\bm{b})\right] 
 &= \sum_{k=1}^{K} \left( \left( ab_kh_k \right)^2\mathbb{E}_{\epsilon}\left[z_k^2(\epsilon T) \right] +1  \right) \nonumber \\
 &\quad - \sum_{k=1}^{K} 2\left( ab_kh_k \right)\mathbb{E}_{\epsilon}\left[z_k(\epsilon T)\right] + \sigma^2a^2 \nonumber \\
    &\quad + \sum_{\substack{q = -  \frac{\mu}{2} \nonumber \\
 q \neq 0}}^{ \frac{\mu}{2} } \sum_{k=1}^{K} \hspace{-0.5mm}\left( ab_kh_{k} \right)^2 \hspace{-0.5mm} \mathbb{E}_{\epsilon}\hspace{-0.5mm}\left[ z_k^2(T(q+\epsilon)) \right] \nonumber \\
 &\!=\! \sum_{k=1}^{K} \left( \left(ab_kh_k \right)^2\tilde{\epsilon}_{0} +1 \right) - 2\sum_{k=1}^{K} ab_kh_k \check{\epsilon} \nonumber \\
 &\quad + \sum_{\substack{q = -  \frac{\mu}{2}  \\
    q \neq 0}}^{ \frac{\mu}{2} } \sum_{k=1}^{K} \left(ab_kh_{k}\tilde{\epsilon}_{q} \right)^2 + \sigma^2a^2 \nonumber \\
&= \mathrm{MSE}^{\mathrm{ISI}}(a,\bm{b}),
\end{align}
which is equal to \eqref{eq:MSE_ISI_q} and proves the equivalence in MSE between considering a single receiver sync error and multiple transmitter errors.

\bibliographystyle{IEEEtran}
\bibliography{OTA_Waveforms.bib}

\begin{IEEEbiography}[{\includegraphics[width=1in,height=1.25in,clip,keepaspectratio]{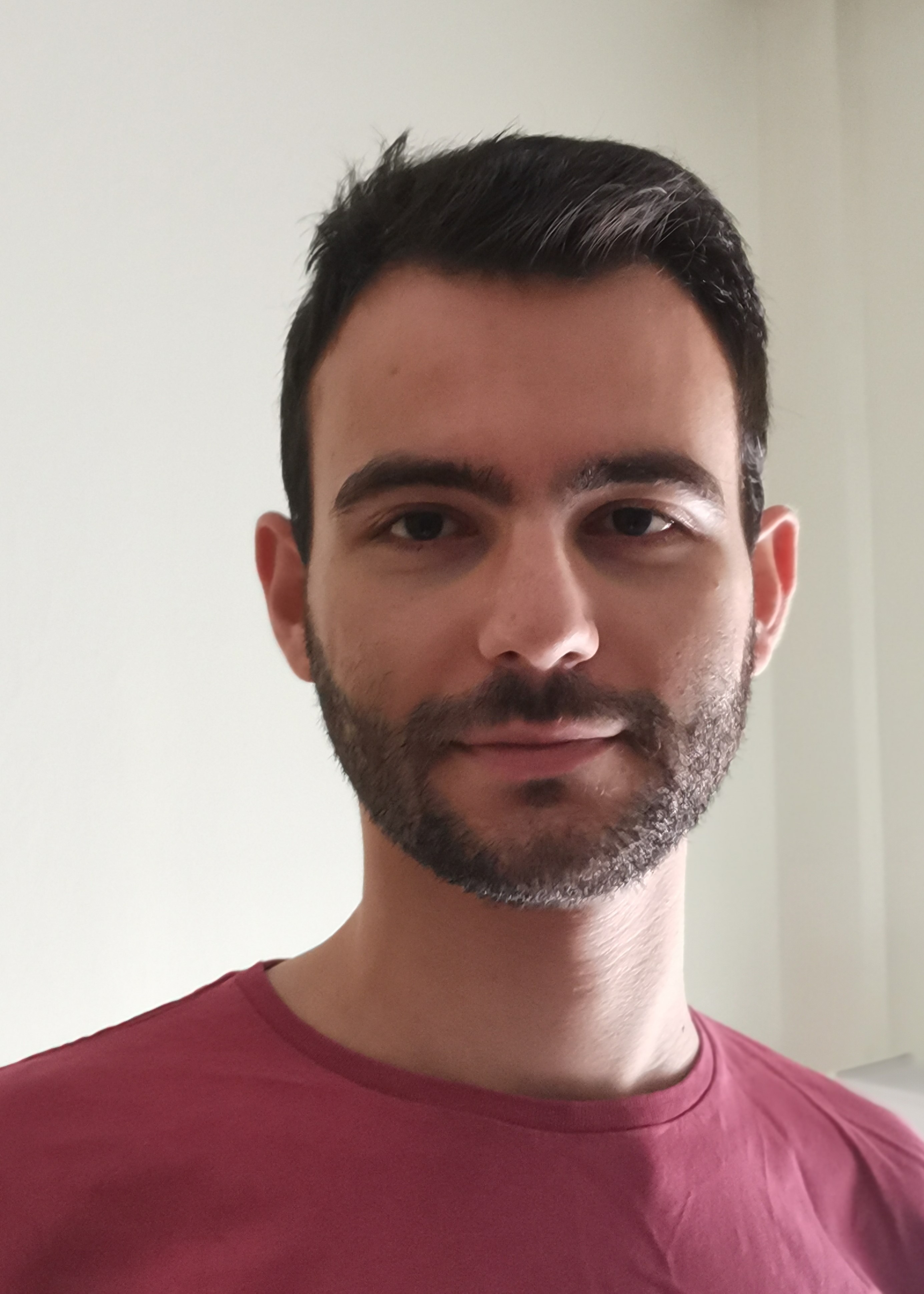}}] {Nikos G. Evgenidis}  received the Diploma (5 years) in Electrical and Computer Engineering from the Aristotle University of Thessaloniki, Greece, in 2022, where he is currently pursuing his PhD. He is also a member of the Wireless and Communications and Information Processing (WCIP) group. He received the Best Paper Award in 2025 Wireless Communications and Networking Conference (WCNC). He was an exemplary reviewer in IEEE Communications Letters in 2024 (top 3\% of reviewers). His major research interests include semantic communications, over-the-air computing, non-orthogonal multiple access, machine learning and optimization theory.
\end{IEEEbiography}

\begin{IEEEbiography}
[{\includegraphics[width=1in,height=1.25in,clip,keepaspectratio]{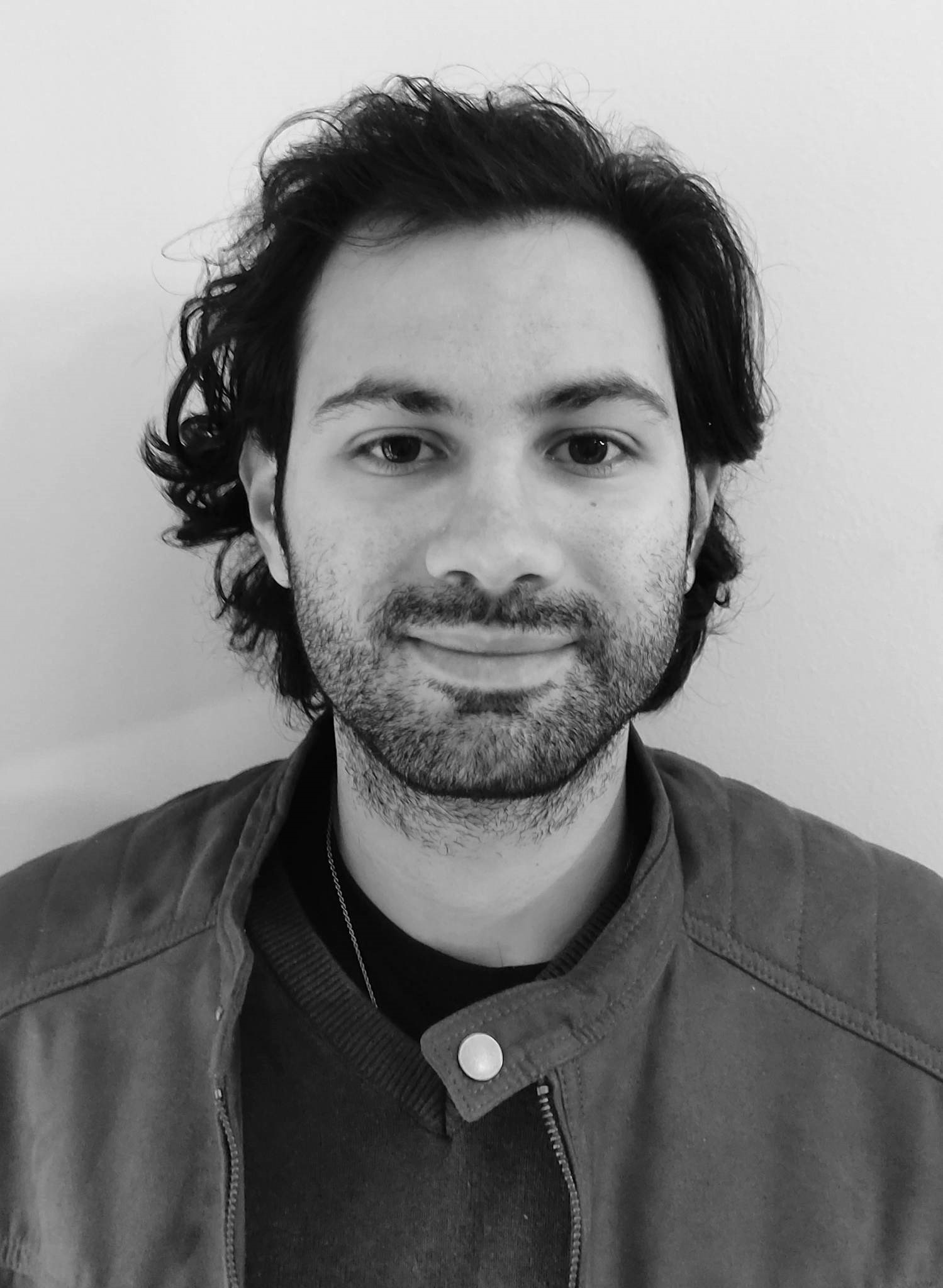}}]{Nikos A. Mitsiou}~(Graduate Student Member, IEEE) received the Diploma Degree (5 years) in Electrical and Computer Engineering from the Aristotle University of Thessaloniki (AUTH), Greece, in 2021, where he is currently pursuing his PhD with the Department of Electrical and Computer Engineering. He is a member of the Wireless and Communications \& Information Processing (WCIP) Group. He was an Exemplary Reviewer of the IEEE Wireless Communications Letters in 2022 (top 3\% of reviewers). He also received the best paper award at IEEE WCNC 2025. His research interests include the intersection of optimization theory and machine learning with applications in wireless networks.
\end{IEEEbiography}

\begin{IEEEbiography}
[{\includegraphics[width=1in,height=1.25in,clip,keepaspectratio]{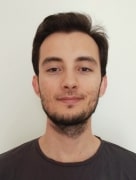}}]
{Sotiris A. Tegos} (Senior Member, IEEE) received the Diploma (5 years) and the Ph.D. degree from the Department of Electrical and Computer Engineering, Aristotle University of Thessaloniki, Thessaloniki, Greece, in 2017 and 2022, respectively. Since 2022, he is a Postdoctoral Fellow with the Wireless Communications and Information Processing (WCIP) Group, Aristotle University of Thessaloniki. In 2018, he was a visitor researcher at the Department of Electrical and Computer Engineering at Khalifa University, Abu Dhabi, UAE. His current research interests include multiple access in wireless communications, wireless power transfer, and optical wireless communications. He serves as an Editor for IEEE Transactions on Communications and IEEE Communications Letters. He received the Best Paper Award in 2023 Photonics Global Conference (PGC) and in 2025 IEEE Wireless Communications and Networking Conference (WCNC). He was an exemplary reviewer in IEEE Wireless Communications Letters in 2019, 2022 and 2023 (top 3\% of reviewers) and an exemplary Editor in IEEE Communications Letters in 2024.
\end{IEEEbiography}

\begin{IEEEbiography}
[{\includegraphics[width=1in,height=1.25in,clip,keepaspectratio]{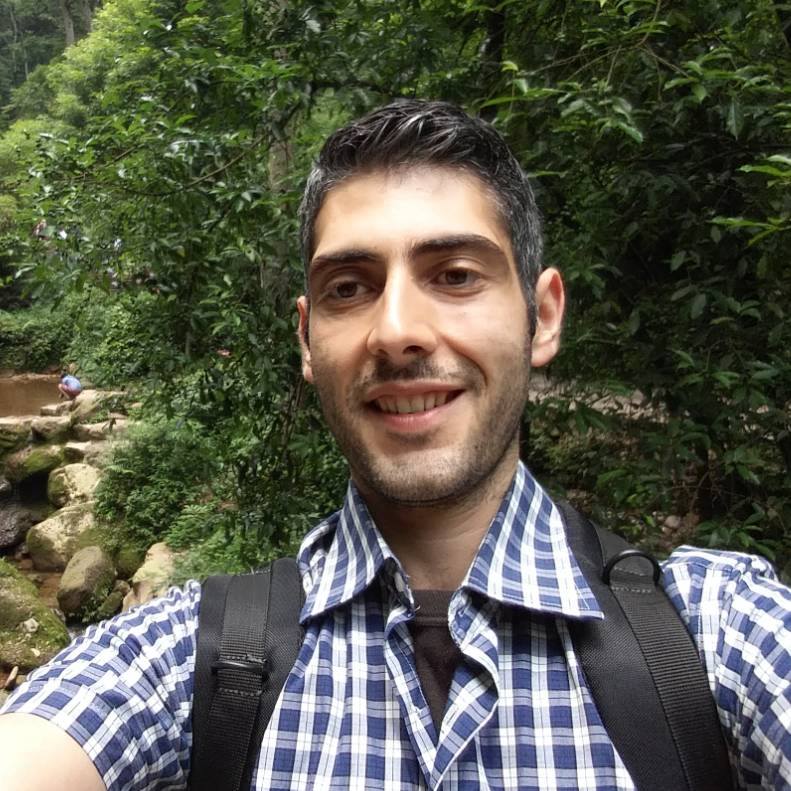}}]
{Panagiotis D. Diamantoulakis} (Senior Member, IEEE) received the Diploma (5 years) and the Ph.D. degree from the Department of Electrical and Computer Engineering, Aristotle University of Thessaloniki, Thessaloniki, Greece, in 2012 and 2017, respectively. From 2017 to 2024, he was a Postdoctoral Fellow with Wireless Communications and Information Processing (WCIP) Group, AUTH and since 2021, he has been a Visiting Assistant Professor with the Key Lab of Information Coding and Transmission, Southwest Jiaotong University, Chengdu, China. In May 2024, he joined the faculty of the Aristotle University of Thessaloniki, Greece, where he is currently Assistant Professor at the Department of Electrical and Computer Engineering.  His research interests include optimization theory and applications in wireless networks, optical wireless communications, and goal-oriented communications. He serves as an Editor of IEEE Open Journal of the Communications Society, while during 2018-2023 he has been an Editor of IEEE Wireless Communications Letters, in which he was an Exemplary Editor of the IEEE Wireless Communications Letters in 2020. Also, he was an Exemplary Reviewer of the IEEE Communications Letters in 2014 and the IEEE Transactions on Communications in 2017 and 2019 (top 3\% of reviewers). 
\end{IEEEbiography}

\begin{IEEEbiography}
[{\includegraphics[width=1in,height=1.25in,clip,keepaspectratio]{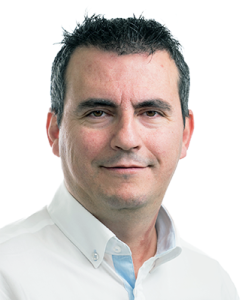}}]
{Panagiotis Sarigiannidis} (Member, IEEE) is the Director of the ITHACA lab (https://ithaca.ece.uowm.gr/), co-founder of the 1st spin-off of the University of Western Macedonia: MetaMind Innovations P.C. (https://metamind.gr) and the I3S spin-off (https://i3slabs.com/en/), and Full Professor in the Department of Electrical and Computer Engineering in the University of Western Macedonia, Kozani, Greece. He received the B.Sc. and Ph.D. degrees in computer science from the Aristotle University of Thessaloniki, Thessaloniki, Greece, in 2001 and 2007, respectively. He has published over 400 papers in international journals, conferences and book chapters, including IEEE Communications Surveys and Tutorials, IEEE Transactions on Communications, IEEE Internet of Things, IEEE Transactions on Broadcasting, IEEE Systems Journal, IEEE Wireless Communications Magazine, IEEE Open Journal of the Communications Society, IEEE/OSA Journal of Lightwave Technology, IEEE Transactions on Industrial Informatics, IEEE Access and Computer Networks. He received 8 best paper and presentation awards and the IEEE SMC TCHS Research and Innovation Award 2023. He has been involved in several national, European and international projects, coordinating and technically leading numerous national and European projects including H2020, Horizon Europe, Erasmus+ and operational programs. His research interests include internet of things, cybersecurity and telecommunications. He is an IEEE member and participates in the Editorial Boards of various journals like IEEE Transactions on Communications, IET Networks, International Journal of Communication Systems and International Journal of Information Security.
\end{IEEEbiography}

\begin{IEEEbiography}
[{\includegraphics[width=1in,height=1.25in,clip,keepaspectratio]{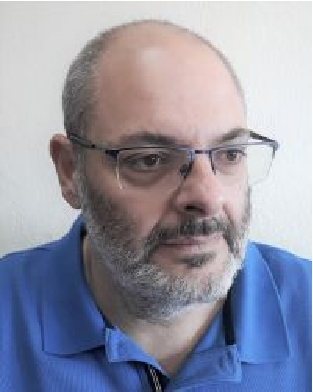}}]
{Ioannis T. Rekanos} (Senior Member, IEEE) received the Diploma degree in electrical engineering and the Ph.D. degree in electrical and computer engineering from the Aristotle University of Thessaloniki (AUTH), Thessaloniki, Greece, in 1993 and 1998, respectively.
From 2000 to 2002, he was a Post-Doctoral Researcher with the Radio Laboratory, Helsinki
University of Technology (now Aalto University), Espoo, Finland. From 2002 to 2006, he was with
the Department of Informatics and Communications, Technological Educational Institute of Central Macedonia, Serres, Greece. From 2006 to 2013, he was with the Physics Division, School of Engineering, AUTH. He is currently a Professor of wave propagation with the School of Electrical and Computer Engineering, AUTH. From 2022 to 2025, he served as Head of the School of Electrical and Computer Engineering, AUTH. Currently, he is Vice-Rector for Research and Innovation at AUTH.
His research interests include electromagnetic and acoustic wave propagation, inverse scattering,
computational electromagnetics and acoustics, and digital signal processing.
\end{IEEEbiography}

\begin{IEEEbiography}
[{\includegraphics[width=1in,height=1.25in,clip,keepaspectratio]{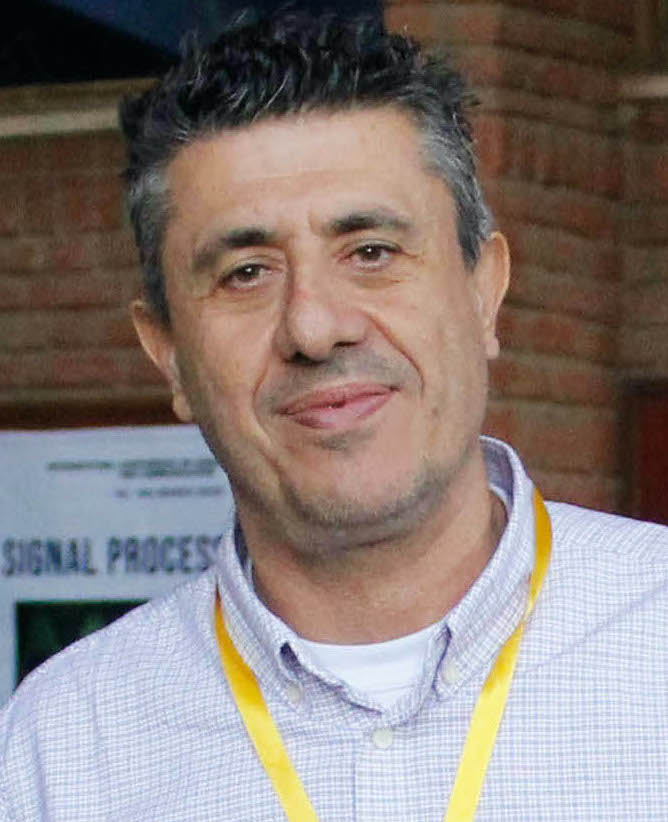}}]
{George K. Karagiannidis} (Fellow, IEEE) received the Ph.D. degree in Telecommunications Engineering from Electrical Engineering Department, University of Patras, Greece, in 1998. He is currently a Professor with the Electrical and Computer Engineering Department, Aristotle University of Thessaloniki, Thessaloniki, Greece, and the Head of Wireless Communications and Information Processing (WCIP) Group. His research interests are in the areas of wireless communications systems and networks, signal processing, optical wireless communications, wireless power transfer, and signal processing for biomedical engineering.
Dr. Karagiannidis recently received three prestigious awards: The 2021 IEEE ComSoc RCC Technical Recognition Award, the 2018 IEEE ComSoc SPCE Technical Recognition Award, and the 2022 Humboldt Research Award from Alexander von Humboldt Foundation. He is one of the Highly Cited Authors across all areas of Electrical Engineering, recognized from Clarivate Analytics as the Web-of-Science Highly-Cited Researcher in the ten consecutive years 2015–2024. Currently, he is the Editor-in Chief of IEEE Transactions on Communications and in the past was the Editor-in Chief of IEEE Communications Letters.
\end{IEEEbiography}

\end{document}